\documentclass[journal]{IEEEtran}
\usepackage{dsfont, subfigure}
\usepackage[english]{babel}
\usepackage{amsthm}
\usepackage{amssymb,mathtools,mathrsfs,bm}
\usepackage{array, multirow}
\usepackage[latin1]{inputenc}
\usepackage{tabularx, graphics, graphicx}
\usepackage{booktabs} 
\usepackage[colorlinks=true,allcolors=cyan]{hyperref}
\usepackage{arydshln}
\usepackage{algorithm}
\usepackage{color}
\usepackage{url}
\usepackage[noadjust]{cite}
\newtheorem{assumption}{Assumption}
\newtheorem{theorem}{Theorem}

\newtheorem{remark}{Remark}
\newtheorem{lemma}{Lemma}
\newtheorem{corollary}{Corollary}

\ifCLASSINFOpdf
\else
\fi
\begin{document}
%
\title{Data-driven Predictive Tracking Control based on Koopman Operators}
%
%

\author{Ye~Wang,
        Yujia~Yang,
        Ye~Pu,
        and~Chris~Manzie 
\thanks{Y. Wang, Y. Yang, Y. Pu and C. Manzie are with the Department of Electrical and Electronic Engineering, The University of Melbourne, Parkville VIC 3010, Australia. {\tt\small  E-mail: \{ye.wang1, ye.pu, manziec\}@unimelb.edu.au, yujyang1@student.unimelb.edu.au}}}
\maketitle

\begin{abstract}
  Constraint handling during tracking operations is at the core of many real-world control implementations and is well understood when dynamic models of the underlying system exist, yet becomes more challenging when data-driven models are used to describe the nonlinear system at hand. We seek to combine the nonlinear modeling capabilities of a wide class of neural networks with the constraint-handling guarantees of model predictive control (MPC) in a rigorous and online computationally tractable framework. The class of networks considered can be captured using Koopman operators, and are integrated into a Koopman-based tracking MPC (KTMPC) for nonlinear systems to track piecewise constant references. The effect of model mismatch between original nonlinear dynamics and its trained Koopman linear model is handled by using a constraint tightening approach in the proposed tracking MPC strategy. By choosing two Lyapunov functions, we prove that solution is recursively feasible and input-to-state stable to a neighborhood of both online and offline optimal reachable steady outputs in the presence of bounded modeling errors under mild assumptions. Finally, we demonstrate the results on a numerical example, before applying the proposed approach to the problem of reference tracking by an autonomous ground vehicle.
\end{abstract}


%
\IEEEpeerreviewmaketitle

\section{Introduction}
%
%
%
%


\IEEEPARstart{W}{ith} the advent of wide scale data collection across increasing application domains, the prospect of data-driven techniques for nonlinear system identification has fast become a reality. At the forefront of this development, neural networks have shown great promise in capturing key features of dynamic system evolution, notably on the back of universal approximation-like properties \cite{kidger2020universal}.


With a background in infinite-dimensional system identification, the Koopman operator \cite{koopman1931hamiltonian} has been attracting attention as a modeling technique for known and unknown nonlinear dynamics, primarily through a process of expressing a nonlinear system as a \emph{lifted} linear model with nonlinear basis functions, see e.g., \cite{brunton2016koopman,otto2021koopman,bevanda2021koopman,kutz2016dynamic}. A challenge in applying the Koopman operator to a physical system is to properly choose the type and number of lifting functions (also called observables or Koopman eigenfunctions) to appropriately lift the  original nonlinear states into a higher dimensional space~\cite{mezic2020koopman}. Recently, deep neural networks have been suggested as the lifting operator with these \emph{deep Koopman} models investigated for general physical systems \cite{lusch2018deep} as well as specific applications including autonomous systems \cite{shi2022deep,xiao2022deep,chen2022offset}.

In the field of control systems, model predictive control (MPC) promises significant advantages over conventional techniques \cite{Maciejowski2002}, \cite{Rawlings2019}. A system operated by an MPC controller can explicitly handle state and input constraints, whilst providing important stability guarantees under reasonable assumptions. Within this context of controlled systems, trajectory tracking represents an important extension in applications from process control to autonomous robots. In \cite{Limon2008}, a tracking MPC framework was proposed for linear system and therefore extended into nonlinear systems \cite{Limon2018}. A robust version of this tracking MPC for linear systems can be also found in~\cite{limon2010robust}. The advantage of this form of tracking MPC is that including a decision variable  of optimal reachable outputs (determined online) improves recursive feasibility and guarantees convergence close to the specified reference. In \cite{berberich2022linear1}, a similar tracking MPC framework using a linearized model of the original nonlinear system was investigated - albeit only for input constraints.

Over the past decade, learning-based control methods have attracted much attention due to the need to provide models that capture individual plant responses and different environmental conditions. Amongst these proposed solutions are a number of investigations in data-driven MPC \cite{korda2018linear,korda2020optimal,wangkoopman,calderon2021koopman,narasingam2022data,zhang2022robust,bruder2020data}. In \cite{korda2018linear,korda2020optimal}, MPC with Koopman linear model was proposed for the regulation problem of possibly unknown nonlinear dynamical systems. In \cite{narasingam2022data}, Koopman operator theory is combined with Lyapunov-based MPC for feedback stabilization of nonlinear systems. In \cite{zhang2022robust}, robust MPC with Koopman operators was studied for a regulation problem, where the control objective is to drive the closed-loop nonlinear system states into a neighbourhood of coordinate origin in the presence of unknown but bounded disturbances. However, these results do not directly translate to tracking problems common in many applied domains. In \cite{bruder2020data}, MPC with Koopman model was applied to soft robots for tracking references. The recent works on learning-based MPC have shown great potential on some stability guarantees for the systems even with unknown dynamics. However, few work on this research line focused on achieving more complex control objectives such as time-varying trajectory tracking. 

The main contribution of this paper is to propose a data-driven tracking MPC based on Koopman operators, referred to hereon as Koopman Tracking MPC (KTMPC), for nonlinear systems with unknown nonlinear dynamics. By choosing a suitable finite-dimensional lifting function, e.g. polynomial, radial basis function or deep neural network, an approximated Koopman linear model for the original nonlinear system can be obtained from  collected data. For general nonlinear systems, modeling errors from a Koopman linear model exist due to the finite dimension of the selected lifted space. To handle modeling errors, an iterative robust constraint tightening approach is applied to  ensure state and input constraint satisfaction, which provides recursive feasibility of nonlinear system with the proposed KTMPC in closed-loop. Using a Lyapunov approach, we prove that the proposed KTMPC guarantees input-to-state stability to a neighborhood of an optimal reachable steady output in closed-loop. 

The remainder of this paper is organized as follows: the problem statement and introduction to Koopman operator theory are discussed in Section~\ref{section:problem statement} along with the link to neural networks. The KTMPC approach is subsequently proposed in Section~\ref{section:MPC and main theorem}. To show the guarantees of the proposed KTMPC, the closed-loop properties are analyzed in details in Section~\ref{section:closed-loop properties}. Results of a numerical example and an application of an autonomous ground vehicle (AGV) are shown in Section \ref{section:examples} to demonstrate the proposed control scheme, before conclusions are drawn in Section \ref{section:conclusion}.

\paragraph*{Notation}
For a matrix $ X $, we use  $\| X \| $and $ X^{\top} $ to denote the induced 2-norm (matrix norm) and the transpose of $ X $, respectively. We use $ X \succ 0 $ to denote a positive definite matrix. We use $ I_n $ to denote an identity matrix of dimension $n$. We also use $\mathbf{0}$ to denote a zero matrix of appropriate dimension. We define the following sets: $ \mathbb{S}^{n} := \left\lbrace X \in \mathbb{R}^{n \times n} : X = X^{\top} \right\rbrace $, $ \mathbb{S}^{n}_{\succ 0} := \{ X \in \mathbb{S}^{n }: X \succ 0 \} $. For any two sets $ \mathcal{X} $ and~$ \mathcal{Y} $, we use the symbols~$ \oplus $ and~$ \ominus $ to denote the Minkowski sum and Pontryagin difference, defined as follows: $ \mathcal{X} \oplus \mathcal{Y} = \left\lbrace x+y : x\in \mathcal{X}, y\in \mathcal{Y} \right\rbrace $, $\mathcal{X} \ominus \mathcal{Y} = \left\lbrace z: z+y \in \mathcal{X}, \forall y \in \mathcal{Y} \right\rbrace $. For a vector~$ x \in \mathbb{R}^{n} $ and~$ P \in \mathbb{S}^{n}_{\succ 0} $, we use~$ \left\| x \right\| $ and $ \left\| x \right\|_P $ to denote the 2-norm and the weighted 2-norm by $P$, respectively. For any two vectors $x \in \mathbb{R}^n$, $y \in \mathbb{R}^n$ with $P \in \mathbb{S}^{n}_{\succ 0}$, the following inequalities hold:
\begin{subequations}
    \begin{align}
        \left \| x + y \right\|  &\leq  \left \| x \right\| + \left \| y \right\| ,\\
        \left \| x + y \right\|_{P}^2  &\leq 2 \left \| x \right\|_{P}^2 + 2\left \| y \right\|_{P}^2, \label{eq:inequality sum norm bound}\\
        \left \| x \right\|_{P}^2 - \left \| y \right\|_{P}^2  &\leq \left \| x - y \right\|_{P}^2 + 2 \left \| x - y \right\|_{P} \left \| y \right\|_{P} \label{eq:inequality two norm minus}.
    \end{align}
\end{subequations}

\section{Problem Statement and Preliminaries}\label{section:problem statement}

Consider the class of nonlinear systems described as
\begin{equation}\label{eq:general nonlinear system}
    x(k+1) = f \left ( x(k),u(k) \right),
\end{equation}
where $x \in \mathbb{R}^{n_x} $ and $ u \in \mathbb{R}^{n_u}$ denote the vectors of states and inputs, respectively. $f \left ( x(k),u(k) \right)$ is a nonlinear function that may be unknown in practice. Furthermore, the system output vector $y \in \mathbb{R}^{n_y}$ is a measurement of some system states and the output equation can be formulated as
\begin{align}
    y(k) = C x(k),
\end{align}
with $C \in \mathbb{R}^{n_y \times n_x}$. Notice that if all system states can be measured as system outputs, then $C = I_{n_x}$. 

The states and inputs of the system \eqref{eq:general nonlinear system} are constrained by
\begin{align}
    x(k) \in \mathcal{X}, u(k) \in \mathcal{U},\; k \in \mathbb{N},
    \label{eq:constraints}
\end{align}
where $\mathcal{X}$ and $\mathcal{U}$ are compact and convex sets, respectively.


\begin{remark}
    A special case of the nonlinear dynamics in \eqref{eq:general nonlinear system}, as discussed in \cite{korda2020optimal}, is 
    \begin{align*}
        x(k+1) = g \left ( x(k) \right) + H u(k),
    \end{align*}
    where $H \in \mathbb{R}^{n_x \times n_u}$, $g (x(k))$ is a nonlinear function. Both of them could be unknown. This special case indicates the independence of control inputs. Since the original nonlinear dynamics are considered to be unknown, we keep the general nonlinear model as in \eqref{eq:general nonlinear system}.
\end{remark}

\subsection{Koopman Operator}\label{subsec:Koopman operator}

Let us first review the Koopman operator theory for discrete-time autonomous dynamic system $x(k+1) = f(x(k),0)$. The Koopman operator $\mathbf{K}: \mathcal{F} \rightarrow \mathcal{F} $ is defined by \cite{korda2018linear}
\begin{equation}\label{eq:Koopman operator}
    \mathbf{K} \bm{\psi}(x) = \bm{\psi}(x) \circ f(x,0),
\end{equation}
where $\circ$ denotes the composition operator. $\bm{\psi}: \mathbb{R}^{n_x} \rightarrow \mathcal{F}$ is the lifting function of $x$, which lifts the original state space $\mathbb{R}^{n_x}$ to a higher dimensional space $\mathcal{F}$. In principle, the lifted space is infinite-dimensional but a possibly finite-dimensional space of the Koopman operator is normally considered. Starting from an initial state $x(0)$, it holds at a time instant $k$ 
\begin{align}
    x(k) = \bm{\psi}^{-1} \left ( \mathbf{K}^k  \bm{\psi}(x(0)) \right ),
\end{align}
where $\bm{\psi}^{-1}: \mathcal{F} \rightarrow \mathbb{R}^{n_x} $ denotes a left-inverse mapping function. 


\subsection{Koopman Linear Model}

For obtaining the Koopman model for the controlled nonlinear system \eqref{eq:general nonlinear system}, the corresponding terms in a finite-dimensional lifted space can be defined as
\begin{align}
    z = \bm{\psi}(x), 
\end{align}
where $ z \in \mathbb{R}^{n_z} $ denotes the vector of the lifted states.  The lifting function $\bm{\psi}(x)$ and the left inverse mapping function $\bm{\psi}^{-1}(z)$ can be chosen by using the extended dynamic mode decomposition (EDMD) method with collected data of system states and inputs \cite{korda2018linear,korda2020optimal,folkestad2020extended}. For complex dynamics, choosing the EDMD modes appropriately can become non-trivial \cite{folkestad2020extended} and so there has been interests in data-driven approaches to replace the need for explicit mode selection. In particular, the use of neural networks to estimate the functions $\bm{\psi}(x)$ and $\bm{\psi}^{-1}(z)$ (respectively referred to as an encoder and a decoder) has been proposed as an implicit approach with the same goal \cite{lusch2018deep}. 

\begin{figure}[thbp]
	\centering
	\includegraphics[width=\columnwidth]{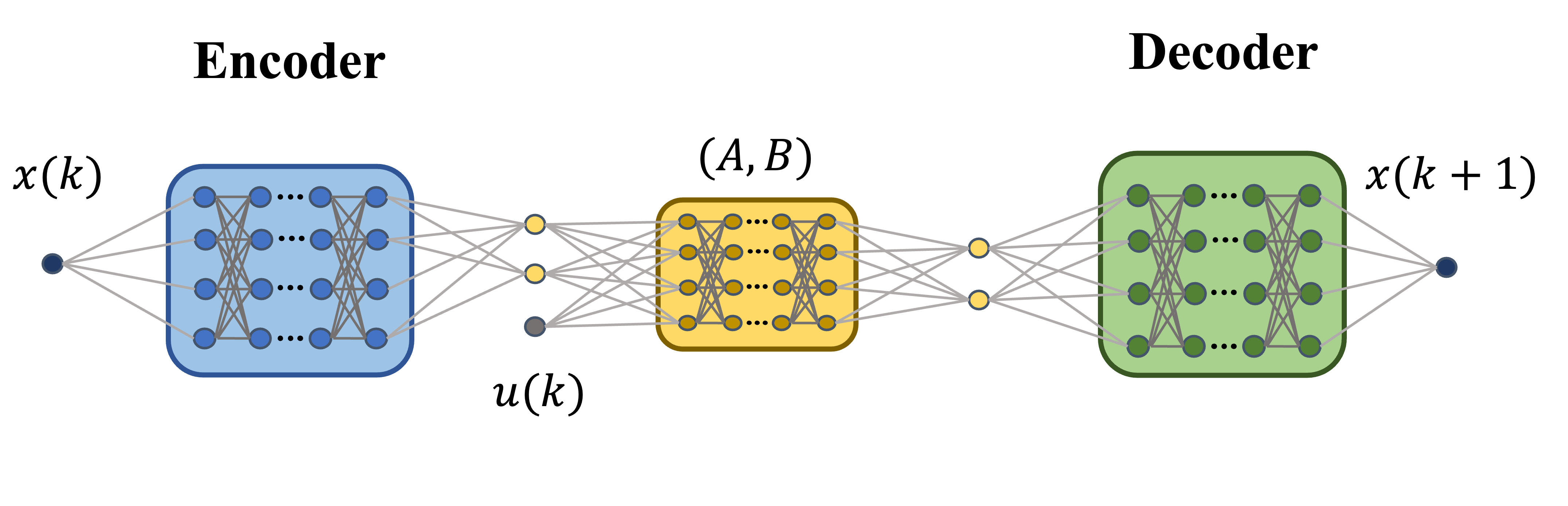}
 \vspace{-0.5cm}
	\caption{General Koopman model based on neural networks.}
	\label{fig:deep koopman}
\end{figure}

A wide range of neural network architectures can be used in this way including radial basis function networks, multi-layer feedforward networks, and convolutional neural networks. The scheme for obtaining a deep Koopman model is described in Figure \ref{fig:deep koopman}. The middle part (in yellow area) only contains the linear layer without bias to impose linear structure of $A$ and $B$. One option to build encoder and decoder is to use standard multi-layer neural networks. Each hidden layer can have the form of $ \mathbf{W} \mathbf{x} + \mathbf{b}$ ($\mathbf{W}$ and $\mathbf{b}$ representing weight and bias) with an activation function, e.g. the rectified linear unit (ReLU). For training the encoder $\bm{\psi}(x)$, the decoder $\bm{\psi}^{-1}(z)$ and the Koopman linear predictor, the explicit loss functions including the reconstruction, the linearity of dynamics, the multi-step prediction, as well as the term with the infinity norm, can be used as introduced \cite{lusch2018deep}.




\subsection{Summary of System Models}

In the following, we introduce several mathematical models for the nonlinear system in \eqref{eq:general nonlinear system}. These models will be used to establish and analyze the tracking MPC with Koopman linear model.

\subsubsection{Equivalent Koopman Model}

For the nonlinear system~\eqref{eq:general nonlinear system}, an equivalent linear predictor based on the Koopman operator is defined as follows:
\begin{subequations}\label{eq:perfect Koopman linear model}
    \begin{align}
        z(k+1) &= A z(k) + B u(k) + w(k; z,x,u),\\
        x(k) &= \bm{\psi}^{-1}(z(k)) + v(k;z),
    \end{align}
\end{subequations}
with $z(0) = \bm{\psi}(x(0))$, where $A \in \mathbb{R}^{n_z \times n_z}$ and $B \in \mathbb{R}^{n_z \times n_u}$ constitute a finite-dimensional approximation of the Koopman operator. $w(k;z,x,u) \in \mathbb{R}^{n_z} $ and $ v(k; z) \in \mathbb{R}^{n_x} $ denote the vectors of disturbances acting in the lifted state $z$ and the original state $x$, respectively.

\begin{assumption}[Disturbance Boundedness]\label{assump:disturbance boundedness}
    In the Koopman model \eqref{eq:perfect Koopman linear model}, the disturbance vectors $w(k;z,x,u)$ and $v(k; z)$ may be unknown but bounded in convex and compact sets:
    \begin{align*}
        w(k;z,x,u) \in \mathcal{W}, v(k;z) \in \mathcal{V}, \; \forall k \in \mathbb{N}, \forall x \in \mathcal{X}, \forall u \in \mathcal{U}.
    \end{align*}
\end{assumption}

\subsubsection{Nominal Koopman Model}

Since the disturbances $w(k;z,x,u)$ and $v(k;z)$ are unknown in practice, we introduce $ \bar{x} \in \mathbb{R}^{n_x} $, $ \bar{z} \in \mathbb{R}^{n_z} $ and $\bar{u} \in \mathbb{R}^{n_u}$ to represent nominal states, nominal lifted state and nominal control inputs. The following nominal Koopman model may be subsequently used in the design of tracking MPC:
\begin{subequations}\label{eq:nominal Koopman linear system}
    \begin{align}
        \bar{z}(k+1) &= A \bar{z}(k) + B \bar{u}(k),\\
        \bar{x}(k) &= \bm{\psi}^{-1}(\bar{z}(k)).\label{eq:linear decoder}
    \end{align}
\end{subequations}

The models in \eqref{eq:perfect Koopman linear model} and \eqref{eq:nominal Koopman linear system} differ through the presence of disturbances from modeling errors. These errors can accumulate if the nominal model is used consecutively as in an MPC formulation. Therefore, in the design of tracking MPC, it is necessary to consider these disturbances to guarantee constraint satisfaction over a prediction horizon, as considered in the next section.

\section{Koopman-based Tracking MPC}\label{section:MPC and main theorem}

In this section, we propose a Koopman-based tracking MPC. The objective is to build a linear optimization formulation for nonlinear system to achieve trajectory tracking while guaranteeing robust constraint satisfaction.

\begin{figure}[thbp]
	\centering
	\includegraphics[width=\columnwidth]{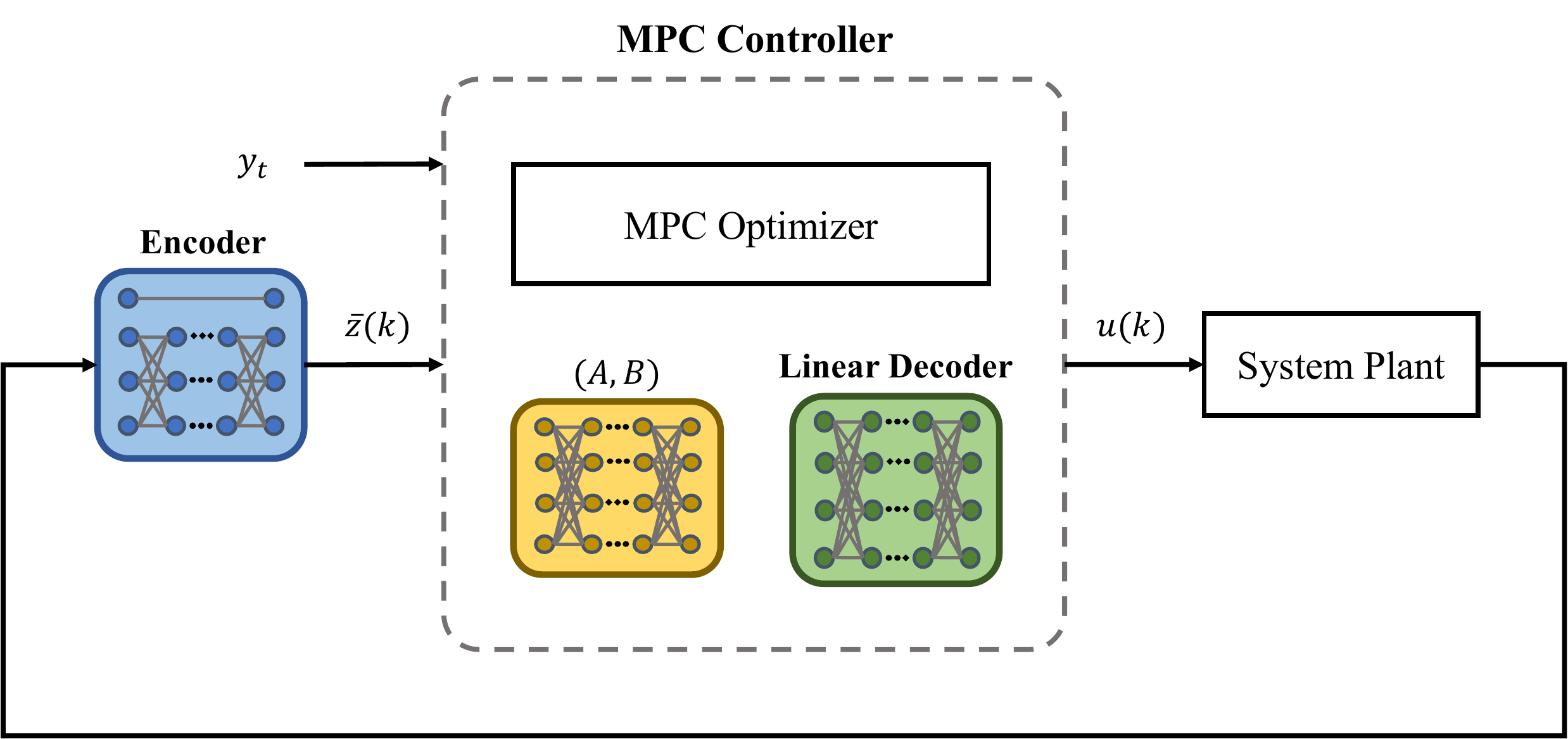}
	\caption{The general scheme of Koopman-based tracking MPC.}
	\label{fig:koopman linear model}
\end{figure}

The general scheme of the proposed Koopman-based tracking MPC is shown in Figure \ref{fig:koopman linear model}. For a nonlinear system~\eqref{eq:general nonlinear system}, the Koopman model is available via offline training with collected data from the original system. The prediction model of the proposed MPC controller is based on the linear Koopman model in lifted space. This prediction model will evaluate the system trajectory along the prediction horizon $N > 0$. To ensure the original nonlinear system \eqref{eq:general nonlinear system} satisfies state and input constraints in \eqref{eq:constraints} while maintaining a linear MPC optimization problem, we impose a decoder of the form:
\begin{align}\label{eq:koopman linear decoder}
    \bm{\psi}^{-1}(\bar{z}) = C_x \bar{z},
\end{align}
where $\bar{z} \in \mathbb{R}^{n_z} $ is a lifted state vector and $C_x \in \mathbb{R}^{n_x \times n_z}$. Therefore, the nominal Koopman linear model in \eqref{eq:nominal Koopman linear system} can be reformulated as 
\begin{subequations}\label{eq:Koopman linear system}
    \begin{align}
        \bar{z}(k+1) &= A \bar{z}(k) + B \bar{u}(k),\\
        \bar{x}(k) &= C_x \bar{z}(k).\label{eq:linear decoder}
    \end{align}
\end{subequations}

\begin{remark}    
    To demonstrate the existence of one possible linear decoder in \eqref{eq:koopman linear decoder}, we consider the encoder with the structure $\bm{\psi}(x) = [x^\top, {\bm{\phi}}(x)^\top]^\top$, where ${\bm{\phi}}(\cdot)$ denotes the additional $(n_z -n_x)$ outputs of the encoder. This leads to $C_x = [I_{n_x} , \mathbf{0} ]$. It is noted however, that other options such as excluding the decoder entirely as in \cite{otto2019linearly}, may be possible if the structure of the state constraints in the original and lifted domains allows these formulations.
\end{remark}

\begin{assumption}\label{assump:controllability}
    For the nonlinear system \eqref{eq:general nonlinear system}, the pair $(A,B)$ in \eqref{eq:Koopman linear system} is controllable (stabilizable). 
\end{assumption}


With the linear decoder in \eqref{eq:koopman linear decoder}, the nominal output equation can be formulated as 
\begin{align}
    \bar{y}(k) &= C_y \bar{z}(k),
\end{align}
where $\bar{y} \in \mathbb{R}^{n_y} $ denotes the nominal output vector and the output matrix can be defined as $C_y := C C_x$.

\subsection{Constraint Tightening Approach}

Due to the inevitable mismatch between the original nonlinear system~\eqref{eq:general nonlinear system} and the Koopman model \eqref{eq:Koopman linear system}, a robust constraint tightening approach \cite{mayne2005robust} is proposed in the MPC problem formulation in order to provide recursive feasibility guarantees.



\begin{remark}
    As in \cite{zhang2022robust}, if the lifting function $\bm{\psi}(x)$ is chosen to be continuous, the set $\mathcal{Z}_{\bm{\psi}} = \lbrace z \in \mathcal{F} \mid z = \bm{\psi}(x), x \in \mathcal{X} \rbrace $ is bounded. Then the boundedness on disturbances $w(k;z,x,u)$ and $v(k;z)$ in Assumption \ref{assump:disturbance boundedness} holds.
\end{remark}


The estimation error between the nonlinear system \eqref{eq:general nonlinear system} and the Koopman model \eqref{eq:Koopman linear system} can be defined as
\begin{align*}
    e(k) := & x(k) - \bar{x}(k)\\
         = & C_x z(k) + v(k;z) - C_x \bar{z}(k)\\
         = & C_x e_z(k) + v(k;z),
\end{align*}
with $e_z(k) := z(k) - \bar{z}(k)$.

By comparing equivalent and nominal Koopman models in~\eqref{eq:perfect Koopman linear model} and \eqref{eq:nominal Koopman linear system}, the relationship of these two control inputs are given in the following.

\begin{assumption}[Local Control Law]\label{assump:local control law}
    For the models \eqref{eq:perfect Koopman linear model} and \eqref{eq:Koopman linear system}, there exists a local control law
    \begin{equation}
        u(k) = \bar{u}(k) + K \big ( z(k) - \bar{z}(k) \big ),
    \end{equation}
    such that $ A_K := A+BK $ is Schur stable, where $K \in \mathbb{R}^{m \times n_z}$.
\end{assumption}

\begin{remark}
    As in \cite{limon2010robust}, this local control law $K$ is required to facilitate constraint satisfaction along the MPC prediction horizon.
\end{remark}

Based on the local control gain $K$, we can obtain the error dynamics for the error $e_z$
\begin{align}\label{eq:e_z dynamics}
    e_z(k+1) = A_K e_z(k) + w(k;z,x,u).
\end{align}

The state and input constraints \eqref{eq:constraints} can be tightened recursively as follows:
\begin{subequations}\label{eq:tightened state and input sets}
    \begin{align}
        \tilde{\mathcal{X}}(j) &:= \mathcal{X} \ominus \left \lbrace C_x \bigoplus_{i=0}^{j-1} (A+BK)^i \mathcal{W} \oplus \mathcal{V} \right \rbrace,\\
        \tilde{\mathcal{U}}(j) &:= \mathcal{U} \ominus \left \lbrace K \bigoplus_{i=0}^{j-1} (A+BK)^i \mathcal{W} \right \rbrace,
    \end{align}
\end{subequations}
for $j = 1,\ldots,N$, with $ \tilde{\mathcal{X}}(0) := \mathcal{X} \ominus \mathcal{V} $, $ \tilde{\mathcal{U}}(0) := \mathcal{U} $.

\begin{assumption}\label{assump:non-empty sets}
    The tightened sets $ \tilde{\mathcal{X}}(N)$ and $\tilde{\mathcal{U}}(N)$ are non-empty.
\end{assumption}

From Assumption \ref{assump:non-empty sets}, it also implies that all the tightened sets $\tilde{\mathcal{X}}(j)$ and $\tilde{\mathcal{U}}(j)$ for $j = 1,\ldots,N$ are non-empty.

\subsection{Steady Manifolds}

In the following, two steady manifolds are introduced for the nonlinear system and its Koopman linear model.

For the nonlinear system \eqref{eq:general nonlinear system}, the steady state and output manifolds are defined as follows:
\begin{align*}
    \mathcal{X}_{u}^s &= \lbrace (x_s,u_s) \mid x_s = f(x_s, u_s), x_s \in \mathcal{X}, u_s \in \mathcal{U} \rbrace,\\
    \mathcal{Y}^s &= \lbrace y_s \in \mathbb{R}^{p} \mid y_s = C x_s,(x_s,u_s) \in \mathcal{X}_{u}^s \rbrace.
\end{align*}

For the Koopman linear system \eqref{eq:Koopman linear system}, the steady state and output manifolds are defined as follows:
\begin{align*}
    \mathcal{Z}_{u}^s(N) &= \{ (z_s,u_s) \mid z_s = A z_s + B u_s, C_x z_s \in \tilde{\mathcal{X}}(N), \\
    & \qquad \qquad \qquad u_s \in \tilde{\mathcal{U}}(N) \},\\
    \tilde{\mathcal{Y}}^s(N) &= \lbrace y_s \in \mathbb{R}^{p} \mid y_s = C_y z_s, (z_s,u_s) \in \mathcal{Z}_{u}^s(N) \rbrace.
\end{align*}

For a given reference point $y_t$, two pairs of steady states, inputs and outputs from the above defined steady manifolds can be obtained by solving the following two optimization problems offline.
\begin{enumerate}
    \item Steady states $x_{sr}$, inputs $u_{sr}$ and outputs $y_{sr}$ of the nonlinear system \eqref{eq:general nonlinear system} can be computed by the optimizer:
    \begin{align}\label{eq:optimizer 1}
      (x_{sr},u_{sr},y_{sr}) := \arg \underset{y_s \in \mathcal{Y}^s}{\mathrm{min}} \left \| y_s - y_{t} \right \|_{S}^2,
    \end{align}
    and the optimal cost of \eqref{eq:optimizer 1} is denoted as $J_{eq}^{*} (y_t)$.
    \item Steady lifted states $\tilde{z}_{sr}$, inputs $\tilde{u}_{sr}$ and outputs $\tilde{y}_{sr}$ of the Koopman linear system \eqref{eq:Koopman linear system} can be computed by the optimizer:
    \begin{align}\label{eq:optimizer sr}
        (\tilde{z}_{sr}, \tilde{u}_{sr}, \tilde{y}_{sr}) := \arg \underset{y_s \in \tilde{\mathcal{Y}}^s(N)}{\mathrm{min}} \left \| y_s - y_{t} \right \|_{S}^2,
    \end{align}
    and the optimal cost of \eqref{eq:optimizer sr} is denoted as $\tilde{J}_{eq}^{*}(y_t)$.
\end{enumerate}

\begin{remark}
    For the above two optimizers, it can be seen that the steady states does not admit $\tilde{z}_{sr} \not = \psi(x_{sr})$.
\end{remark}

\subsection{Koopman-based Tracking MPC Formulation}

Over a prediction horizon $N > 0 $, the KTMPC cost function can be defined as follows:
\begin{subequations}
    \begin{align}
        &J_{N}(x(k),y_t) =  \left \| y_s - y_{t} \right \|_{S}^2 + \sum_{j=0}^{N-1} \ell \left ( \bar{z}(j),\bar{u}(j),z_s,u_s \right ),\label{eq:KTMPC total cost function}\\
        &\ell \left ( \bar{z}(j),\bar{u}(j),z_s,u_s \right ) = \left \|  \bar{z}(j)-z_s \right \|_{Q}^2 + \left \| \bar{u}(j)-u_s \right \|_{R}^2, \label{eq:KTMPC stage cost function}
    \end{align}
\end{subequations}
where $Q \in \mathbb{S}^{n_x}_{\succ 0}$, $R \in \mathbb{S}^{n_u}_{\succ 0}$, and $S \in \mathbb{S}^{n_y}_{\succ 0}$. $y_s$ and $u_s$ are online decision variables representing optimal reachable steady outputs and inputs together with $ \bar{\mathbf{u}} = \{\bar{u}(j), j = 0, \ldots, N-1 \}$ as the MPC decision variables. 

In general, the KTMPC optimization problem can be formulated as follows:
\begin{subequations}\label{problem:KTMPC}
	\begin{align}
	& \underset{\bar{\mathbf{u}},u_s,y_s} {\mathrm{minimize}}\;\; J_{N}(x(k),y_t), \\
	\intertext{subject to}
	& \quad \bar{z}(0) = \bm{\psi}(x(k)),\label{eq:KTMPC initialization}\\
	& \quad \bar{z}(j+1) = A \bar{z}(j) + B \bar{u}(j) , \; j = 0,\ldots, N-1,\label{eq:KTMPC prediction model}\\
	& \quad C_x \bar{z}(j) \in \tilde{\mathcal{X}}(j),\; \bar{u}(j) \in \tilde{\mathcal{U}}(j), \; j = 0,\ldots, N-1,\label{eq:KTMPC state-input constraint}\\
	& \quad z_s = A z_s + B u_s,\label{eq:KTMPC steady states}\\
	& \quad y_s = C_y z_s, \label{eq:KTMPC steady output}\\
	& \quad C_x z_s \in \tilde{\mathcal{X}}(N),\;u_s \in \tilde{\mathcal{U}}(N),\label{eq:KTMPC steady constraint}\\
	& \quad \bar{z}(N) = z_s, \label{eq:KTMPC terminal condition}
	\end{align}
\end{subequations}
where the sets $\tilde{\mathcal{X}}(j)$ and $\tilde{\mathcal{U}}(j)$, $j = 0,\ldots, N-1$ are defined as in \eqref{eq:tightened state and input sets}. The constraints in \eqref{problem:KTMPC} are explained as follows: \eqref{eq:KTMPC initialization} is used to initialize the first predicted state $\bar{z}(0)$, where $x(k)$ is the measured state of nonlinear system~\eqref{eq:general nonlinear system} at every time $k$; \eqref{eq:KTMPC prediction model} is the nominal Koopman linear model, which is used to compute the predicted state in the lifted space; \eqref{eq:KTMPC state-input constraint} includes tightened state and input constraints; \eqref{eq:KTMPC steady states} and \eqref{eq:KTMPC steady output} are the steady state and output equations at the lifted space based on the Koopman linear model;  \eqref{eq:KTMPC steady constraint} guarantees the steady states and inputs satisfy the tightened state and input constraints; \eqref{eq:KTMPC terminal condition} is the terminal equality constraint to ensure that the terminal state has reached the admissible steady state~$z_s$.

At each time step $k \in \mathbb{N}$, solving the above optimization problem can obtain the optimal solutions of the optimization problem \eqref{problem:KTMPC}, denoted by $ \bar{\mathbf{u}}^*(k) = \left[ \bar{u}^*(0 ; k), \ldots, \bar{u}^*(N-1 ; k) \right]^{\top} $, $u^{*}_{s}(k)$ and $y^{*}_{s}(k)$. The optimal cost of KTMPC is denoted by $J_{N}^{*}(x(k),y_t)$. From the optimal solutions, we can also construct the optimal sequence $ \bar{\mathbf{z}}^*(k) = \left[ \bar{z}^*(0 ; k), \ldots, \bar{z}^*(N ; k)\right]^{\top} $. Based on the optimal solution obtained at time step $k$, the optimal control action, applied to the system \eqref{eq:general nonlinear system} at the current time step $k$, can be chosen as
\begin{align}\label{eq:robust control action}
    u(k) = \bar{u}^*(0 ; k).
\end{align}


\section{Analysis of Closed-loop Properties for KTMPC}\label{section:closed-loop properties}

In this section, we discuss recursive feasibility and closed-loop convergence of the nonlinear system \eqref{eq:general nonlinear system} controlled by the KTMPC controller implemented in \eqref{problem:KTMPC}. By using the proposed constraint tightening approach, recursive feasibility of the closed-loop system is guaranteed. To prove closed-loop convergence, we notably define two Lyapunov functions. With these two Lyapunov functions, we address the closed-loop input-to-state stability. The following additional assumptions are necessary for the closed-loop property analysis.

\begin{assumption}[Offset Function \cite{Limon2018}]\label{assump:offset function}
    For the offset function and a steady output $y_s$, there exists a scalar $\alpha_O > 0 $ such that
    \begin{align}\label{eq:ys-ysr inequality}
        \left \| y_s - y_{t} \right \|_{S}^2 - \left \| \tilde{y}_{sr} - y_{t} \right \|_{S}^2 \geq \alpha_O \left \| y_s - \tilde{y}_{sr} \right \|^2.
    \end{align}
\end{assumption}

\begin{assumption}[Uniqueness of Steady States]\label{assump:uniqueness}
    For the Koopman linear model, there exists a unique pair of steady state and input $(z_s,u_s)$.
\end{assumption}

\begin{assumption}[Weak Controllability \cite{Rawlings2019}]\label{assump:weak controllability}
    For a steady pair $(z_s,u_s)$, there exists a scalar $\beta_u >0$ such that
    \begin{align}\label{eq:weak controllability}
        \sum_{j=0}^{N-1} \ell \left ( \bar{z}(j)-z_s, \bar{u}(j)-u_s \right ) \leq \beta_u \left \| \bar{z}(0) - z_{s} \right \|^{2}. 
    \end{align}
\end{assumption}

\begin{assumption}[Local Control Law for Tracking]\label{assump:local control law for tracking}
    Given two sets $\bar{\mathcal{X}}$ and $\bar{\mathcal{U}}$, and a constant reference signal $ y_r \in \mathbb{R}^{n_y} $. For the Koopman linear model \eqref{eq:Koopman linear system}, there exists a local control law $\kappa_f (z(k), y_r)$ for tracking $y_r$ as
    \begin{align}\label{eq:local control law for tracking}
        u(k) = K_z \left( z(k) - z_r \right) + u_r,
    \end{align}
    where $ z(k) = \bm{\psi}(x(k)) $, $K_z \in \mathbb{R}^{n_u \times n_z} $ and the pair $(z_r,u_r)$ corresponds to the reference $y_r$. 
\end{assumption}

\begin{assumption}\label{assump:beta u bound}
    For an optimal steady pair $(z_s^*(k),u_s^*(k))$ obtained from the optimization problem \eqref{problem:KTMPC} at a time step $k$, the following condition holds for the scalar $\beta_u$ in \eqref{eq:weak controllability}:
    \begin{align}\label{eq:beta u1 condition}
        \beta_{u} \leq \frac{1+\sqrt{5}}{2} \underline{\lambda}(Q).
    \end{align}
\end{assumption}

\begin{remark}
    Assumptions \ref{assump:offset function}-\ref{assump:local control law for tracking} are typical conditions for guaranteeing the closed-loop convergence for tracking MPC \cite{limon2010robust,Limon2018,Rawlings2019}. Assumption \ref{assump:beta u bound} introduces an additional condition beyond these standard requirements. One interpretation of Assumption \ref{assump:beta u bound} is it places a requirement on the minimum rate of convergence of the controlled system towards $\tilde{y}_{sr}$, which can be obtained through the appropriate selection of weighting matrices $Q$ and $R$ providing a sufficiently aggressive control action.
\end{remark}

\subsection{Lyapunov Candidate Functions}

In the following, we introduce two Lyapunov candidate functions. According to the optimization problem \eqref{problem:KTMPC}, for a given reference point $y_t$, optimal reachable steady output $y_s^*(k)$ is determined online at each time step $k$ with the MPC implementation while an offline-determined optimal reachable steady output $\tilde{y}_{sr}$ can also be determined by the optimizer \eqref{eq:optimizer sr}. We use the first Lyapunov function to discuss the closed-loop stability with respect to $y_s^*(k)$. Based on this result, we use the second Lyapunov function to develop the closed-loop stability guarantee with respect to $\tilde{y}_{sr}$.

The first Lyapunov candidate function is chosen as
\begin{align}\label{eq:Lyapunov candidate function}
    V_1(x(k),y_t) := \sum_{j=0}^{N-1} \ell \left ( \bar{z}(j;k)-z_s^*(k), \bar{u}(j;k)-u_s^*(k) \right ) .
\end{align}

\begin{lemma}
    Consider Assumptions \ref{assump:offset function}-\ref{assump:weak controllability} hold. For any compact set $\mathcal{Z} \in \mathbb{R}^{n_z}$, there exist two positive scalars $\beta_{l_1} > 0 $ and $ \beta_{u_1} > 0 $ such that
    \begin{align}\label{eq:V1 bounds}
        \beta_{l_1} \left \| z(k) - z_s^* (k) \right\|^2 \leq V_1(x(k),y_t) \leq \beta_{u_1} \left \| z(k) - z_s^* (k) \right\|^2,
    \end{align}
    for all $z(k) = \bm{\psi} (x(k)) \in \mathcal{Z}$.
\end{lemma}

\begin{proof}
    (i) \textbf{Lower Bound:} Based on the definition in \eqref{eq:Lyapunov candidate function}, it is straightforward to obtain the lower bound: 
    \begin{align*}
        V_1(x(k),y_t) 
        & \geq \left \| z(k) - z_s^* (k) \right\|_Q^2 \\
        &\geq \beta_{l_1} \left \| z(k) - z_s^* (k) \right\|^2,
    \end{align*}
    where $\beta_{l_1} := \underline{\lambda}(Q)$.
    
    (ii) \textbf{Upper Bound:}
    The upper bound is given by the weak controllability from Assumption \ref{assump:weak controllability}. For the quadratic function $V_1(x(k),y_t)$, there exists a scalar $\beta_u>0$ such that
    \begin{align*}
        V_1(x(k),y_t)
        &\overset{\eqref{eq:weak controllability}}{\leq}  \beta_{u_1} \left \| z(k) - z_s^* (k) \right\|^2,
    \end{align*}
    which gives \eqref{eq:V1 bounds} together with the lower bound obtained above.
\end{proof}

Taking into account the steady pair and the optimal cost of the optimizer \eqref{eq:optimizer sr}, the second Lyapunov candidate function can be chosen to be
\begin{align}\label{eq:Lyapunov candidate function 2}
    V_2(x(k),y_t) := J_{N}^{*}(x(k),y_t) - \tilde{J}_{eq}^{*}(y_t).
\end{align}

\begin{lemma}\label{lemma:bounds for V2}
    Consider Assumptions \ref{assump:offset function}-\ref{assump:weak controllability} hold. For any compact set $\mathcal{Z} \in \mathbb{R}^{n_z}$, there exist two positive scalars $ \beta_{l_2}> 0 $ and $ \beta_{u_2} > 0 $ such that
    \begin{align}\label{eq:V2 bounds}
        \beta_{l_2} \left \| z(k) - \tilde{z}_{sr} \right\|^2 \leq V_2(x(k),y_t) \leq {\beta_{u_2}} \left \| z(k) - \tilde{z}_{sr} \right\|^2,
    \end{align}
    for all $z(k) = \bm{\psi} (x(k)) \in \mathcal{Z}$.
\end{lemma}

\begin{proof}
    (i) \textbf{Lower Bound:} Considering Assumption \ref{assump:offset function} holds, from the optimization problem \eqref{problem:KTMPC}, we have
    \begin{align*}
        V_2(x(k),y_t) &= J_{N}^{*}(x(k),y_t) - \tilde{J}_{eq}^{*}(y_t)\\
        & \geq \left \| z(k) - z_s^* (k) \right\|_Q^2 + \left \| y_s^*(k) - y_t \right\|_S^2 - \tilde{J}_{eq}^{*}(y_t)\\
        & \overset{\eqref{eq:ys-ysr inequality}}{\geq} \left \| z(k) - z_s^* (k) \right\|_Q^2 + \alpha_O \left \| y_s^*(k) - \tilde{y}_{sr} \right \|^2.
    \end{align*}
    
    Considering Assumption \ref{assump:uniqueness} holds, the steady pair is unique, which implies that there exists a scalar $c_l > 0 $ such that
    \begin{align}\label{eq:ys lower bound}
        \left \| y_s^*(k) - \tilde{y}_{sr} \right \|^2 \geq \frac{1}{c_l} \left \| z_s^*(k) - \tilde{z}_{sr} \right \|^2.
    \end{align}
    
    Then, by combining the above two conditions, we can obtain
    \begin{align*}
        V_2(x(k),y_t) &\geq \underline{\lambda}(Q) \left \| z(k) - z_s^* (k) \right\|^2 + \frac{\alpha_O}{c_l} \left \| z_s^*(k) - \tilde{z} _{sr} \right \|^2 \\
        &\geq \beta_{l_2} \left \| z(k) - \tilde{z}_{sr} \right\|^2,
    \end{align*}
    where $ \beta_{l_2} := \frac{\min \lbrace \underline{\lambda}(Q), \frac{\alpha_O}{c_l}\rbrace}{2} $.
    
    (ii) \textbf{Upper Bound:} Let us consider the pair $(\tilde{z}_{sr}, \tilde{u}_{sr}, \tilde{y}_{sr})$ be the steady pair at a time step $k$, that is, $z_s(k) = \tilde{z}_{sr}$, $u_s(k) = \tilde{u}_{sr} $ and $ y_s(k) = \tilde{y}_{sr}$. There exists a sequence of input $\mathbf{u}(k)$ such that the optimization problem \eqref{problem:KTMPC} is feasible and the corresponding cost is
    \begin{align*}
        \tilde{J}_{N}(x(k),y_t) = \left \| \tilde{y}_{sr} - y_{t} \right \|_{S}^2 +  \sum_{j=0}^{N-1} \ell \left ( z(j)-\tilde{z}_{sr}, u(j)-\tilde{u}_{sr} \right ).
    \end{align*}
    
    Since the optimal cost of \eqref{problem:KTMPC} at time step $k$ is $J_{N}^{*}(x(k),y_t)$, we know
    \begin{align}\label{eq:feasible cost}
        J_{N}^{*}(x(k),y_t) \leq \tilde{J}_{N}(x(k),y_t).
    \end{align}
    
    Considering Assumption \ref{assump:weak controllability} holds, it implies that there exists a scalar $ \beta_{u_2} > 0$ such that
    \begin{align}\label{eq:ineq weak controllability}
        \sum_{j=0}^{N-1} \left \| z(j) -  \tilde{z}_{sr} \right\|_Q^2 + \left \| u(j)- \tilde{u}_{sr} \right\|_R^2 \leq \beta_{u_2} \left \| z(k) - \tilde{z}_{sr} \right\|^2.
    \end{align}
    
    Then, we can have 
    \begin{align*}
        V_2(x(k),y_t) \overset{\eqref{eq:feasible cost}}{\leq} & \tilde{J}_{N}(x(k),y_t)- \tilde{J}_{eq}^{*}(y_t)\\
        \overset{\eqref{eq:ineq weak controllability}}{\leq} & \left \| \tilde{y}_{sr} - y_{t} \right \|_{S}^2 + \beta_{u_2} \left \| z(k) - \tilde{z}_{sr} \right\|^2  - \tilde{J}_{eq}^{*}(y_t)\\
        \leq & \;\beta_{u_2} \left \| z(k) - \tilde{z}_{sr} \right\|^2.
    \end{align*}
    
    Thus, the Lyapunov candidate function $V_2(x(k),y_t)$ defined in \eqref{eq:Lyapunov candidate function} is bounded by two $\mathcal{K}_{\infty}$ functions.
\end{proof}

\subsection{Theoretical Results}

We first discuss recursive feasibility of the nonlinear system \eqref{eq:general nonlinear system} with the proposed KTMPC controller in \eqref{problem:KTMPC}.

\begin{theorem}[Recursive Feasibility]\label{theorem:RF}
    Consider Assumptions \ref{assump:controllability}-\ref{assump:local control law} and \ref{assump:local control law for tracking} hold. The nonlinear system \eqref{eq:general nonlinear system} with the KTMPC controller defined in \eqref{problem:KTMPC} is recursively feasible, for time-varying references $y_t(k)$ at any time step $k \in
    \mathbb{N}$.
\end{theorem}



\begin{proof}
    The proof can be found in Appendix \ref{appendix:theorem RF}.
\end{proof}


We next discuss the closed-loop stability for the nonlinear system \eqref{eq:general nonlinear system} with the KTMPC \eqref{problem:KTMPC}. The stability result is addressed based on the above defined two Lyapunov candidate functions. The following lemma is necessary to introduce for the stability result.

\begin{lemma}\label{lemma:ys bounds}
    Given a reference $y_t$ and consider the optimal reachable steady output $\tilde{y}_{sr}$ and the optimal steady output $y_s^*(k)$ of the KTMPC \eqref{problem:KTMPC}. For any steady output $  y_s = \sigma y_s^*(k) + (1-\sigma) \tilde{y}_{sr}$ with $ \sigma \in [0,1]$, it holds
    \begin{align}\label{eq:y_s(k+1)-ys(k)-yt upper bound}
        \left \| y_s - y_{t} \right \|_{S}^2 &- \left \| y_s^*(k) - y_{t} \right \|_{S}^2 \nonumber\\
         \leq  & -s(2\sigma-\sigma^2) \left \| y_s^*(k) - \tilde{y}_{sr} \right \|^2,
    \end{align}
    where $S = s I$ is chosen to be a scaled identity matrix with a scalar $s > 0$.
\end{lemma}

\begin{proof}
    The proof can be found in Appendix \ref{appendix:lemma 3}.
\end{proof}

\begin{theorem}[Input-to-State Stability]\label{theorem:ISS stability}
    Consider Assumptions \ref{assump:controllability}-\ref{assump:beta u bound} hold. Given the weighting matrices $Q$ and $R$ in \eqref{eq:KTMPC stage cost function}, for a piecewise constant reference point $y_t$, the closed-loop system output $ y(k) $ of the nonlinear system~\eqref{eq:general nonlinear system} with the KTMPC controller defined in~\eqref{problem:KTMPC} is input-to-state stable, i.e. converging to a neighborhood of the optimal reachable steady output $\tilde{y}_{sr}$ obtained from the optimizer \eqref{eq:optimizer sr}, and the size of the neighborhood is determined by the amplitude of estimation errors.
\end{theorem}

\begin{proof}
    The proof can be found in Appendix \ref{appendix:theorem-stability}.
\end{proof}



\begin{remark}
Theorem \ref{theorem:ISS stability} provides implicit design guidelines for neural networks used in the KTMPC scheme, as the tracking error for a piecewise constant reference is directly related to the estimation error of the Koopman model. If certain classes of neural networks are considered (e.g. radial basis functions), inserting additional nodes in the network can provide better approximation properties \cite{kidger2020universal} and hence better tracking results.
\end{remark}

\begin{remark}
    Theorem \ref{theorem:ISS stability} requires a piecewise constant $y_t$ in order to provide input-to-state stability, i.e. we can guarantee a neighbourhood to which the tracking error converges for a constant $y_t$. This requirement is not restrictive in many applications such as mobile robotics or chemical processing where the reference is held constant for consecutive sampling instants.
\end{remark}

From Theorem \ref{theorem:ISS stability}, it can be seen that the closed-loop system is converging to a neighborhood of $\tilde{y}_{sr}$. If we have a perfect Koopman model, then we can have asymptotic stability in the following corollary.

\begin{corollary}\label{corollary:AS stability}
    Consider Assumptions \ref{assump:controllability}-\ref{assump:local control law for tracking} hold and a perfect Koopman model \eqref{eq:perfect Koopman linear model} is available for \eqref{eq:general nonlinear system} such that $w(k;z,x,u) = 0$ and $v(k;z) = 0$, $\forall k \in \mathbb{N}$. The closed-loop system output $ y(k) $ of \eqref{eq:general nonlinear system} with the KTMPC controller in~\eqref{problem:KTMPC} is asymptotically converging to the optimal reachable steady output $\tilde{y}_{sr}$.
\end{corollary}

\begin{proof}
    The proof is analogous to the proof in \cite{Limon2008}. Also, based on the proof of Theorem \ref{theorem:ISS stability}, the class $\mathcal{K}_{\infty}$ functions $\gamma(w)$ and $\tilde{\gamma}(w)$ approach zero as disturbances reduced to zero.
\end{proof}



\section{Simulations and Experiment}\label{section:examples}

\subsection{Numerical Example}\label{subsection:numerical example}



Motivated by \cite{brunton2016koopman}, we first demonstrate the proposed approach  without  neural networks playing the role of encoder and decoder in order to verify the theoretical results related to KTMPC. As a special case, the following nonlinear dynamics:
\begin{align*}
    x_1(k+1) &= \lambda x_1(k),\\
    x_2(k+1) &= \mu x_2(k) + (\lambda^2 - \mu) x_1^2(k) + u(k),
\end{align*}
where parameters $\lambda = -0.1$ and $\mu = 2$ are considered. By selecting the lifting function with polynomial terms as $ \bm{\psi}(x) = \left[x_1(k),x_2(k),x_1^2(k)\right]^\top$, we can obtain the perfect Koopman model in $\mathbb{R}^3$, i.e. there is no error introduced by using a lifted Koopman model. 

To test the proposed KTMPC with scalable disturbances, we augment the Koopman model with additive disturbances $w$ and $v$ as follows:
\begin{align*}
    z(k+1) &=
    \begin{bmatrix}
      \lambda & 0 & 0 \\
      0 & \mu & (\lambda^2 - \mu)\\
      0 & 0 & \lambda^2
    \end{bmatrix} z(k)+ 
    \begin{bmatrix}
      0 \\ 1 \\ 0
    \end{bmatrix} u(k) + w(k),\\
    x(k) &= C_x z(k) + v(k),\\
    y(k) &= C x(k),
\end{align*}
where $w(k) \in \mathcal{W} = \{w \in \mathbb{R}^3 | -0.2 I_3 \leq w \leq 0.2 I_3 \} $, $v \in \mathcal{V} = \{ v \in \mathbb{R}^2 | -0.1 I_2 \leq v \leq 0.1 I_2 \}$, $\forall k \in \mathbb{N}$, are randomly sampled. The system matrices $C_x = [I_2, 0]$ and $C = [0,1,0]$, and the control objective is to track given piecewise constant references $y_t$. 

\begin{figure}[thbp]
	\centering
	\subfigure[$y(k)$]{\includegraphics[width=\hsize]{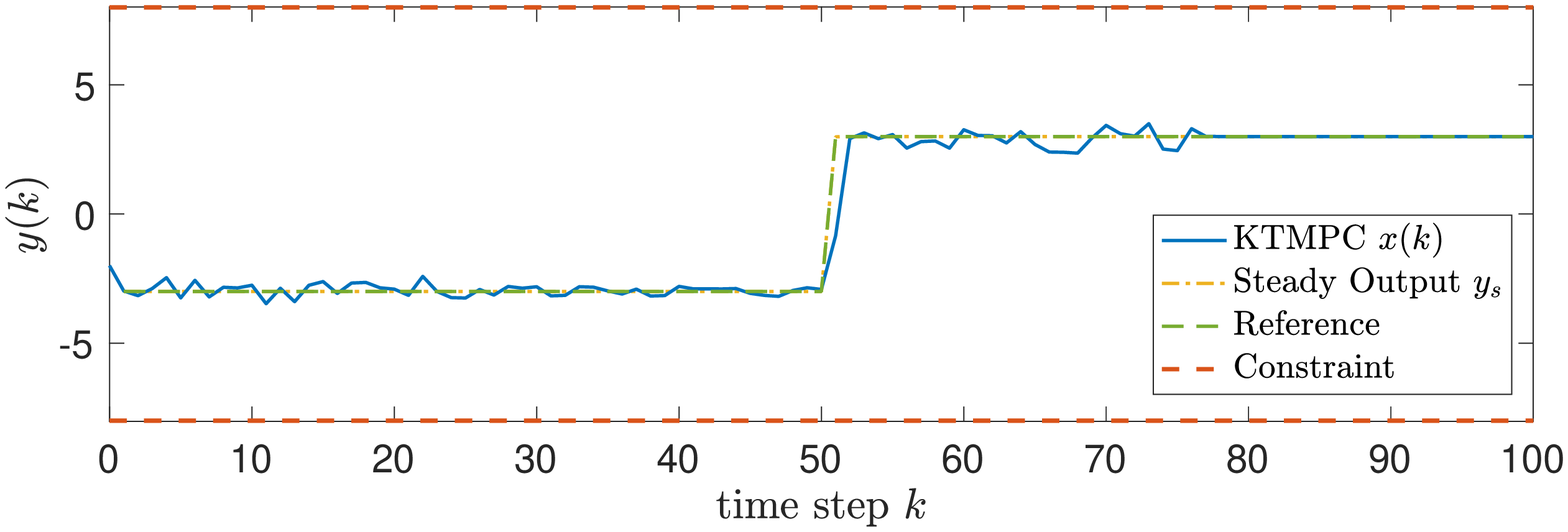}}
	\subfigure[$u(k)$]{\includegraphics[width=\hsize]{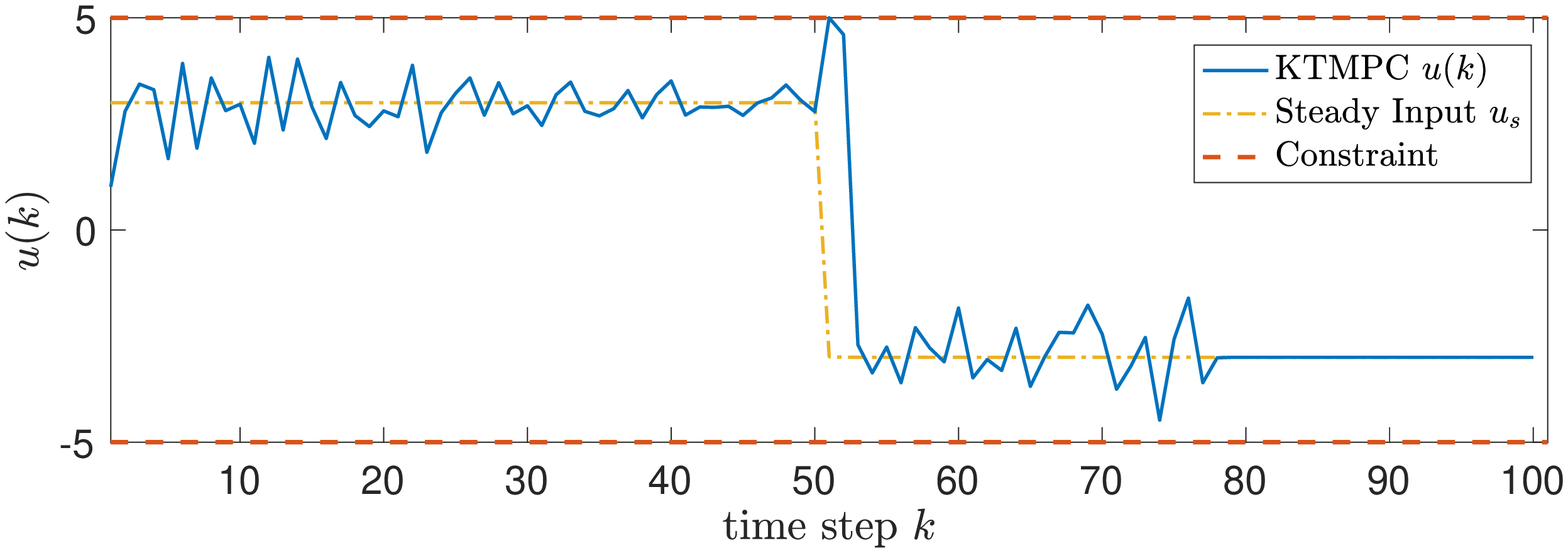}}
	\subfigure[$w(k)$]{\includegraphics[width=\hsize]{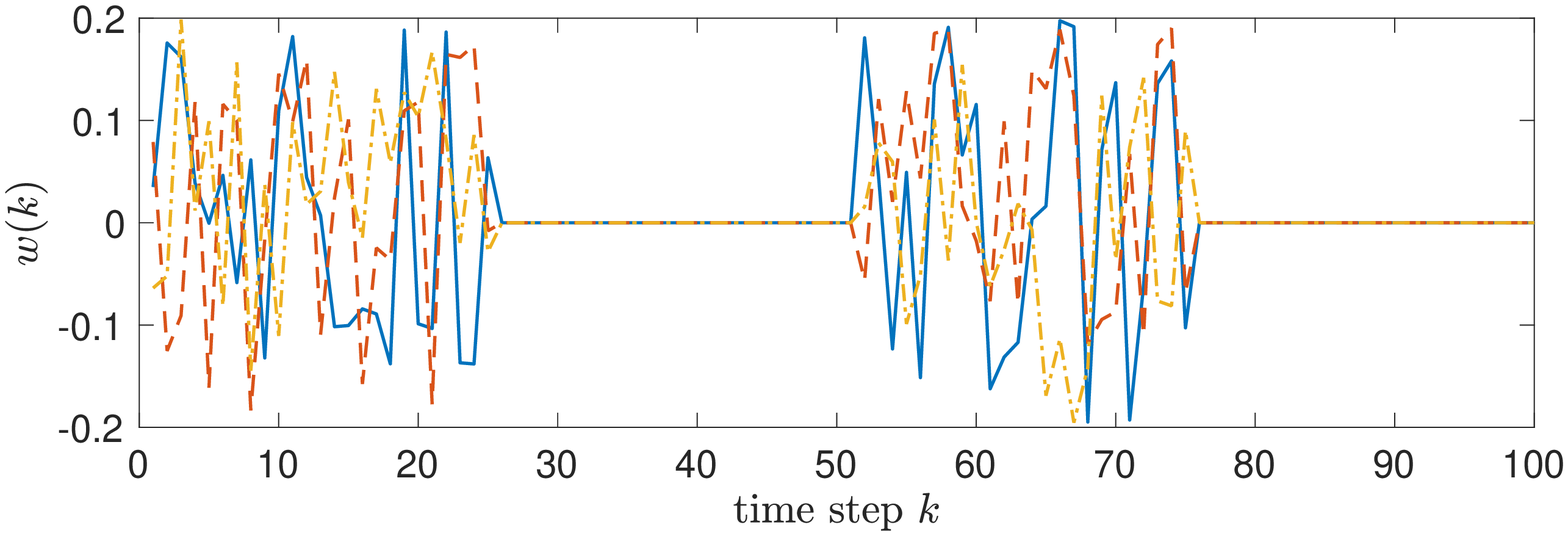}}
	\subfigure[$v(k)$]{\includegraphics[width=\hsize]{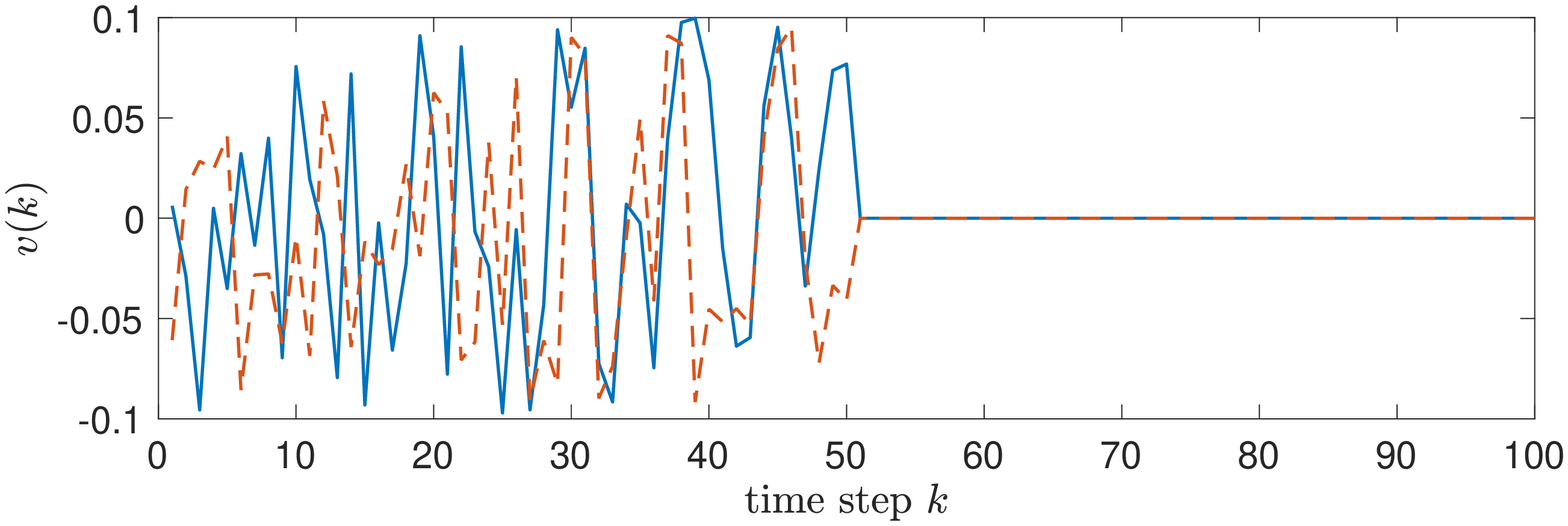}}
	\caption{Simulation results for numerical example in Section \ref{subsection:numerical example}.}
	\label{fig:numerical example}
\end{figure}

As shown in Figure \ref{fig:numerical example}, the closed-loop output converges to a neighborhood of the reference $y_t$ when there exist disturbances $w(k)$ and $v(k)$, thereby providing some validations of Theorem \ref{theorem:ISS stability}. Furthermore, when $w(k)= 0$ and $v(k)= 0$, the closed-loop tracking is perfect, as predicted by Corollary \ref{corollary:AS stability}. In all cases the closed-loop system is shown to satisfy input and output constraints, validating Theorem \ref{theorem:RF} in the presence and absence of estimation errors. We next consider a more realistic example where a perfect Koopman model does not exist.

\subsection{Simulation: an Autonomous Ground Vehicle}

\begin{figure*}[t]
	\centering
	\subfigure[]{\includegraphics[width=0.325\hsize]{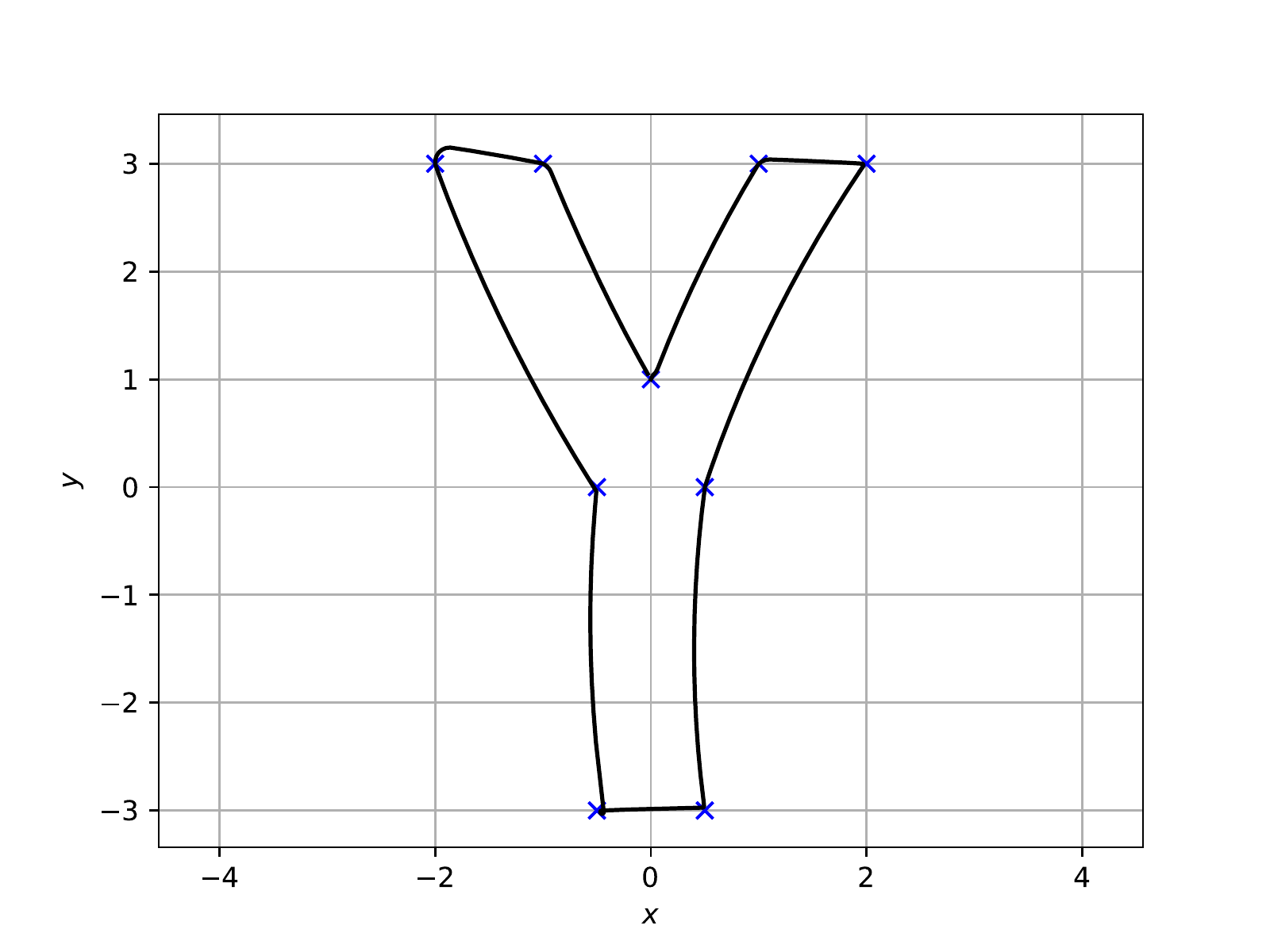}}
	\subfigure[]{\includegraphics[width=0.325\hsize]{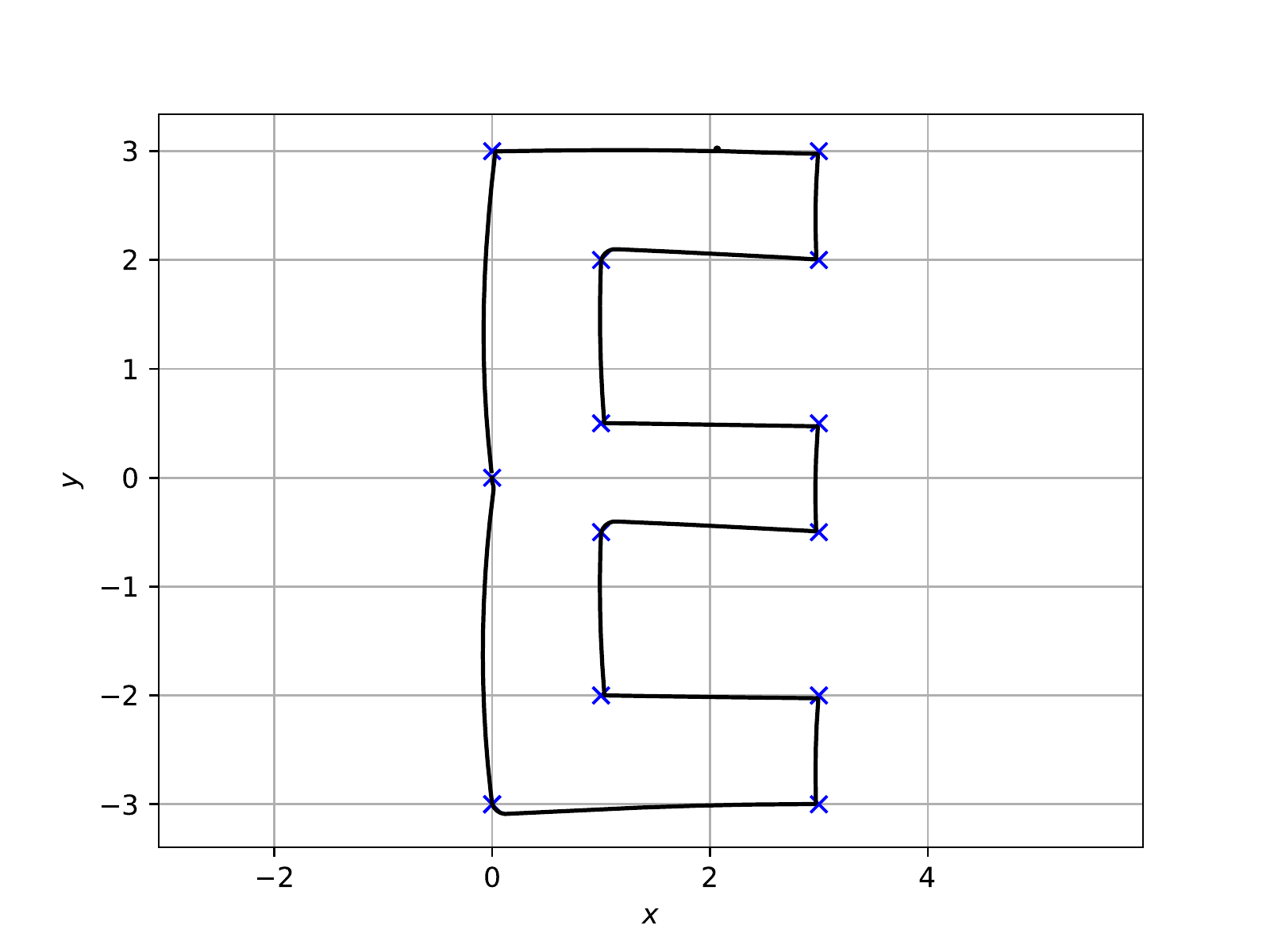}}
	\subfigure[]{\includegraphics[width=0.325\hsize]{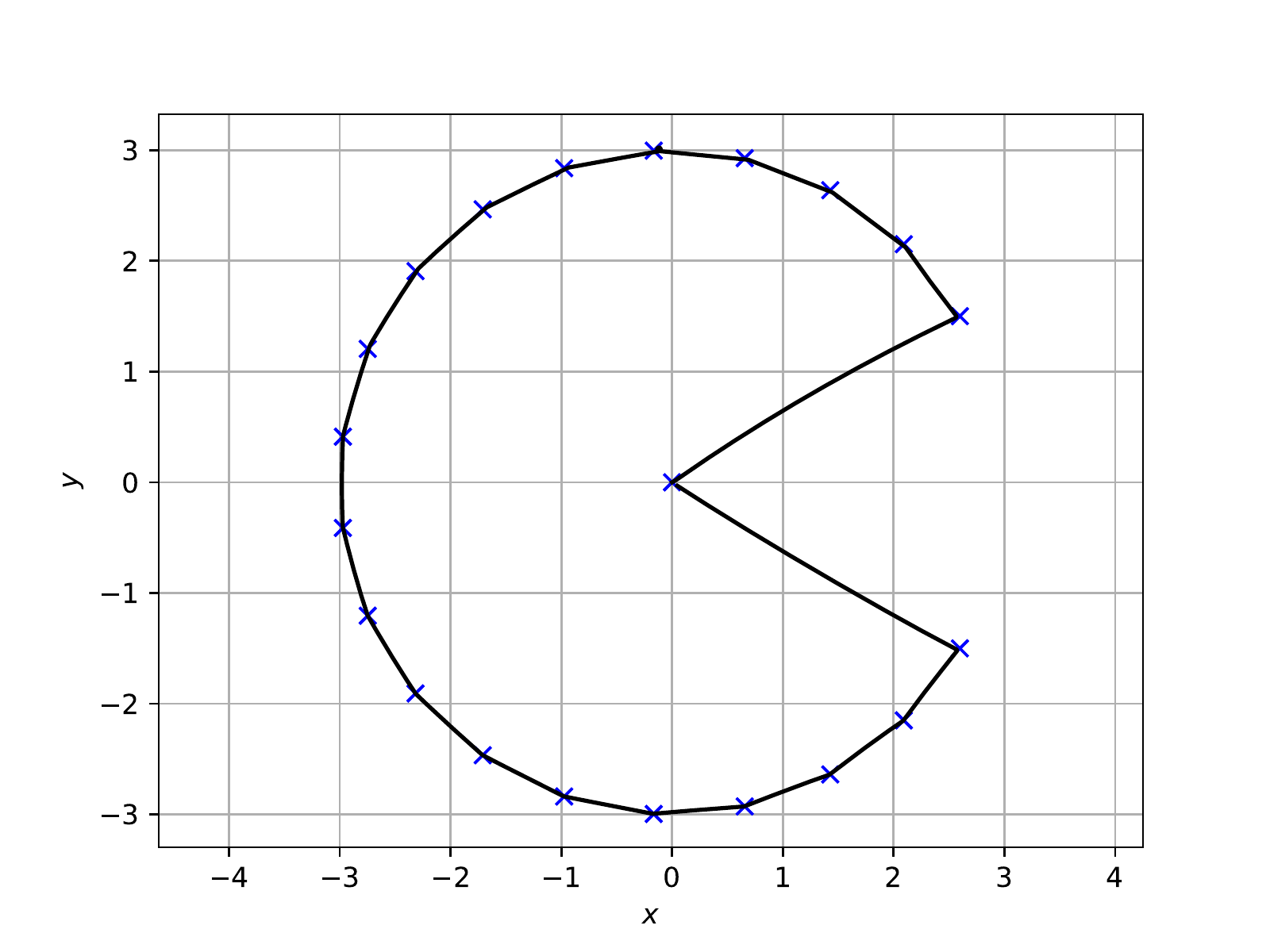}\label{fig:pacman single}}
	\caption{AGV simulation results (piecewise constant references are given in blue cross point, and the closed-loop trajectories are in black solid lines).}
	\label{fig:AGVtotalsim}
\end{figure*}

\begin{figure}[thbp]
	\centering
	\includegraphics[width=\hsize]{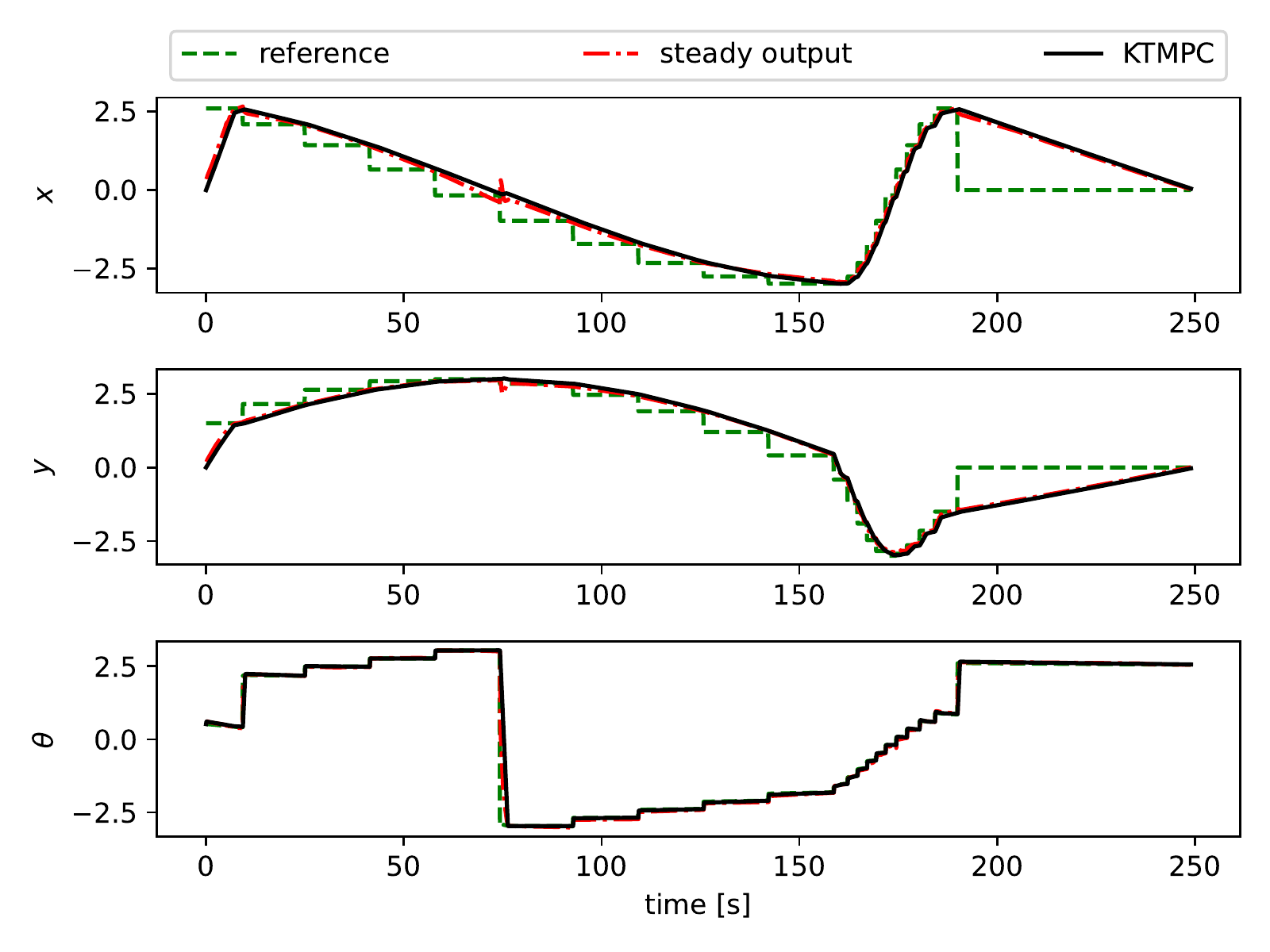}
	\caption{AGV state trajectories using the reference in Figure \ref{fig:pacman single}.}
	\label{fig:AGV_x}
\end{figure}

\begin{figure}[thbp]
	\centering
	\includegraphics[width=\hsize]{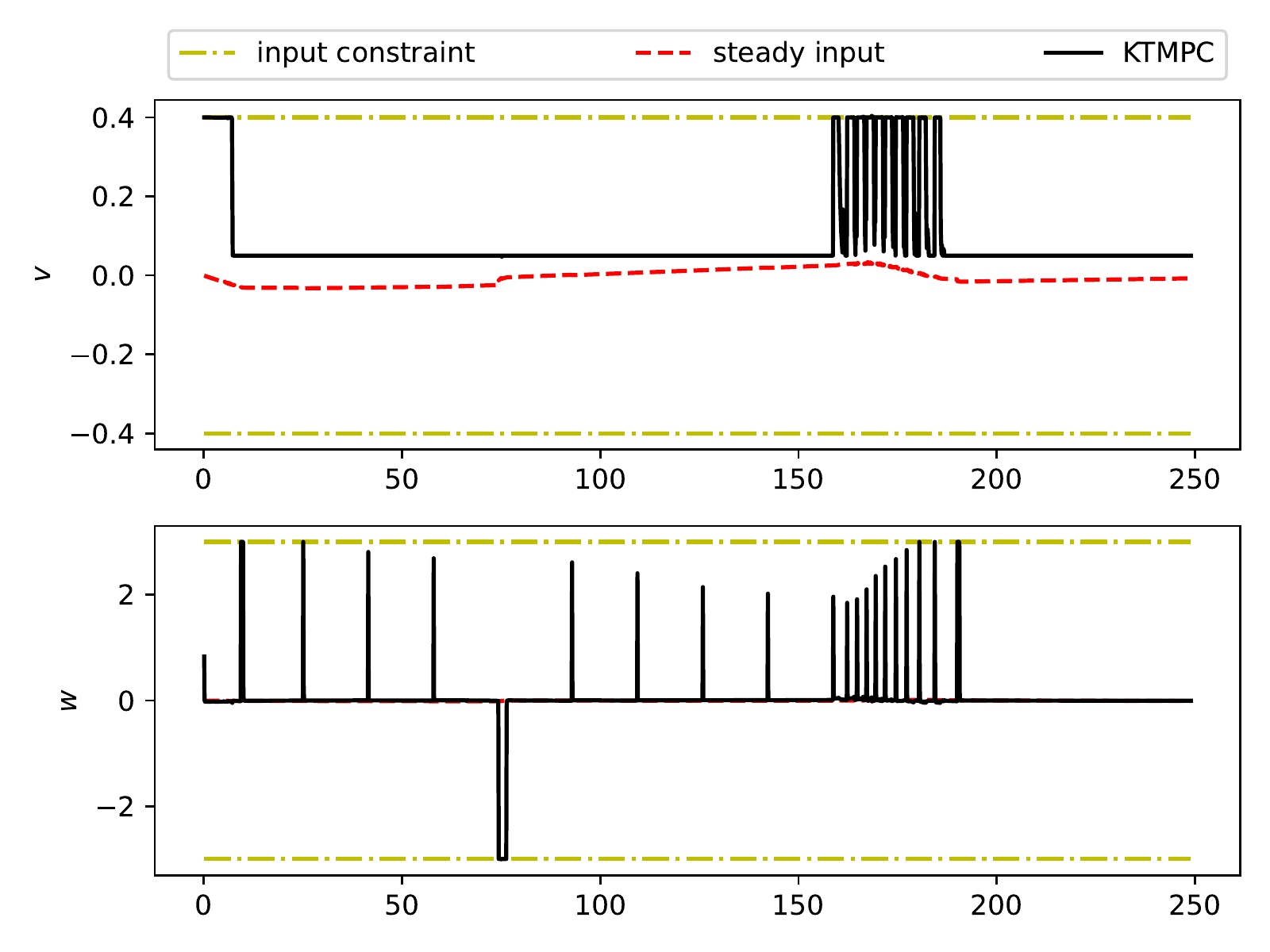}
	\caption{AGV inputs using the reference in Figure \ref{fig:pacman single}.}
	\label{fig:AGV_u}
\end{figure}

We apply the proposed KTMPC controller to an AGV with a nonlinear kinematic unicycle model \cite{zhang2021trajectory}. This application does not require the full generality of the theoretical results in Section \ref{section:closed-loop properties}, but provides a practical demonstration to clearly highlight the implications for even more general systems.

The AGV position $[p_x,p_y]^\top$ and the heading angle $\theta$ are chosen to be system states $x$ while the linear velocity $v_r$ and angular velocity $w_r$ are chosen to be the control inputs $u $. We use a standard multi-layer neural network structure with the ReLU activation function for both encoder and decoder. Specifically, we choose the dimension of the lifted space to be $n_z = 11$. The original inputs are kept in the Koopman model such that the input constraints still remain linear. The training data were collected by using the unicycle AGV model. Specifically, 10,0000 trajectories with a simulation length of 10 time steps (sampling time of 0.1 s) were collected. Although the considered unicycle model with chosen states and inputs is a control-affine nonlinear system, it is still possible to obtain a Koopman model as described in \eqref{eq:perfect Koopman linear model} with unknown disturbances $w(k;z,x,u)$ and $v(k;z)$. After obtaining the deep Koopman model, the disturbance sets were evaluated by a sample-based method. The trained deep Koopman AGV model and the simulation codes are available via the link\footnote{\url{https://github.com/autosysproj/KTMPC.git}}.

The simulation results with three reference trajectories are shown in Figure \ref{fig:AGVtotalsim}. From three simulation results, it can be seen that the AGV can well track the given piecewise constant references. Regarding the reference in Figure \ref{fig:pacman single}, the corresponding states and inputs are shown in Figure \ref{fig:AGV_x} and Figure \ref{fig:AGV_u}, respectively. As shown in Figure \ref{fig:AGV_x}, the references are given in green dashed lines and the optimal reachable steady outputs shown in red dashed and dotted lines are obtained online from the KTMPC optimization. The closed-loop states with the KTMPC in black solid lines are converging to both references and steady outputs. As also shown in Figure \ref{fig:AGV_u}, the control inputs obtained by the KTMPC are converging to the steady inputs.

\subsection{Experiment: an Autonomous Ground Vehicle}

\begin{figure}[t]
	\centering
        \includegraphics[width=\hsize]{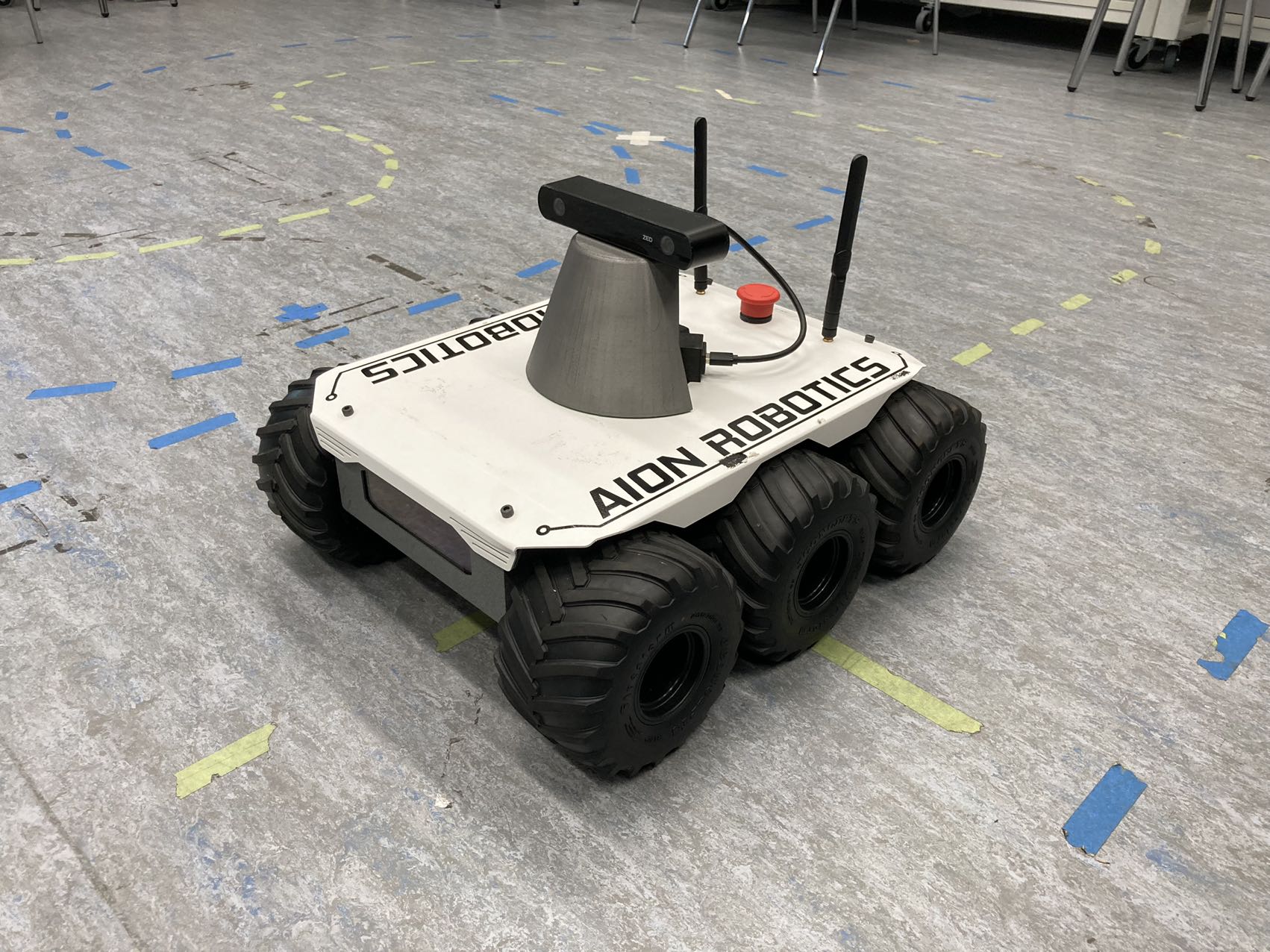}
	\caption{Aion R6 rover with an onboard camera.}
	\label{fig:AGV}
\end{figure}

We next implement the proposed KTMPC with the deep Koopman model obtained from the previous simulation on the Aion R6 AGV (shown in Figure~\ref{fig:AGV}) \cite{aion} to track piece-wise constant reference points sampled along the centerline of a racing circuit. Whilst in the previous subsection, the electric motors were unmodeled, here a hierarchical control approach is used in the experiments. The proposed KTMPC is set as the outer loop controller to provide optimal control actions for linear velocity and the heading angle at each time step while the default inner loop controller in the Aion R6 AGV platform provides the electric motor inputs that track these setpoints. The outer loop controller is run with a sampling rate of 0.1 seconds, which is well within the computational capability required to implement KTMPC on the embedded platform.

\begin{figure}[t]
	\centering
	\includegraphics[width=\hsize]{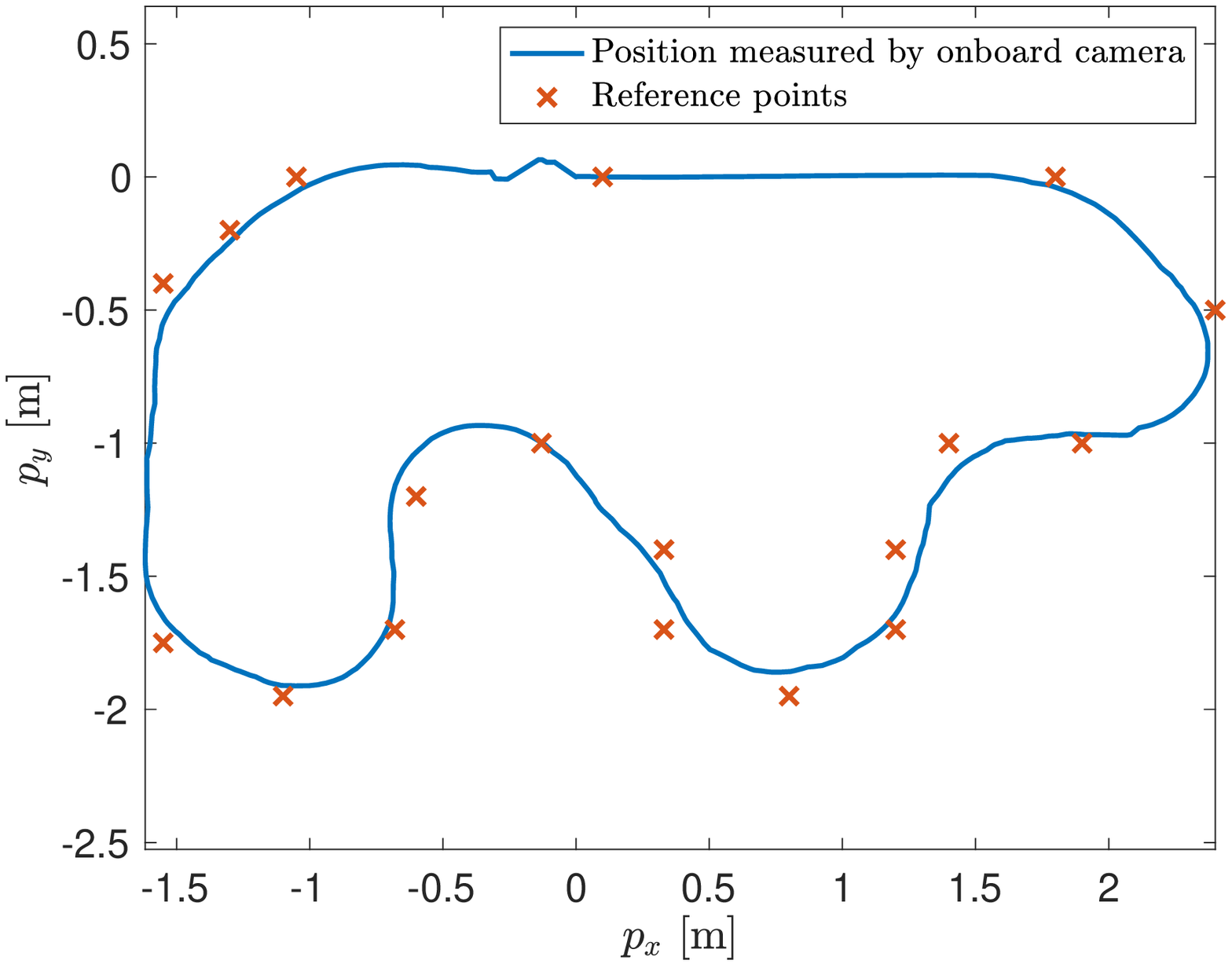}
	\caption{AGV experiment piecewise constant references and a trajectory from the data obtained by the onboard camera and a video recording of the experiment available in \protect\url{https://www.youtube.com/watch?v=iLzOD6_5tuE}.}
	\label{fig:AGV exp}
\end{figure}

An onboard ZED camera is used in the experiment to feed back the real-time position of the AGV to the KTMPC controller. A sequence of references were taken from the circuit. In the experiment, the criterion for switching reference points is when the distance between the reference point and the current position is below 0.35 m. A closed-loop AGV position trajectory is shown in Figure~\ref{fig:AGV exp}. A video recording of this experiment is also available through the link. From the experiment result, it can be seen that the Aion R6 with the KTMPC controller can navigate around the circuit and track the given reference points.

\section{Conclusions}\label{section:conclusion}



In this paper, we have proposed a Koopman-based tracking MPC for nonlinear systems to track piecewise constant references. The nonlinear dynamics are not necessarily known. With collected data from the real system, a Koopman linear model can be obtained by properly choosing approximated lifting functions in a finite-dimensional space with preset neural networks. The Koopman linear model is used as the prediction model in the tracking MPC formulation. To handle modeling errors and guarantee recursive feasibility of MPC, a robust constraint tightening approach is employed. We have proved that although the modeling errors exist, the original nonlinear system with the proposed KTMPC controller is recursively feasible and converging to a neighborhood of optimal reachable output close to a given reference point. Through simulations and an AGV experiment, we have demonstrated the efficacy of the proposed KTMPC. As one future research, controllability conditions will be enhanced when training the Koopman model. Future work may also include investigating the approximation capabilities of different network architectures and sizes, and their impact in different application settings.

\section*{Acknowledgments}

We would like to acknowledge Yiping Qian for training the deep Koopman model of AGV and Xiangyi Zheng for setting up the AGV experiment platform. Ye Wang and Ye Pu acknowledge support from the Australian Research Council via the Discovery Early Career Researcher Awards (DE220100609 and DE220101527), respectively.

\appendices
\section{Proof of Theorem \ref{theorem:RF}}\label{appendix:theorem RF}

This proof can be split into two cases: the one is that the reference is not changed in two consecutive time steps and another is that the reference is changed.

\textbf{(i) With the same reference} $y_t(k+1) = y_t(k)$: At a time step $k \geq 0 $, the optimization problem \eqref{problem:KTMPC} is feasible and the optimal solutions at time step $k$ are
\begin{align*}
   \bar{\mathbf{u}}^*(k) &= \left[ \bar{u}^*(0 ; k), \ldots, \bar{u}^*(N-1 ; k) \right]^{\top},\\
   \bar{\mathbf{z}}^*(k) &= \left[ \bar{z}^*(0 ; k), \ldots, \bar{z}^*(N ; k)\right]^{\top},
\end{align*}
and $ z^{*}_{s}(k)$, $u^{*}_{s}(k)$ and $y^{*}_{s}(k)$. The optimal control action $ u(k) $ can be found .

At the next time step $k+1$, $x(k+1) = f \left ( x(k),u(k) \right)$.

Then, we can define the input sequence at time step $k+1$ as follows:
\begin{align*}
    \bar{u}(j;k+1) &= K (\bar{z}(j;k+1)-\bar{z}^*(j+1;k)) \\
    & \quad + \bar{u}^*(j+1 ; k), \, j = 0, \ldots, N-2,\\
    \bar{u}(N-1;k+1) &= \kappa_f \left ( \bar{z}^*(N; k), y_s^*(k) \right)\\
    &= K_z (\bar{z}^*(N; k)-z_s^*(k)) + u_s^*(k)\\
    & \overset{\eqref{eq:KTMPC terminal condition}}{=}u_s^*(k),
\end{align*}
where $K$ and $K_f$ are from Assumptions \ref{assump:local control law} and \ref{assump:local control law for tracking}. In addition, the associated state sequence at time step $k+1$ are
\begin{align*}
    \bar{z}(0 ; k+1), \ldots, \bar{z}(N-1 ; k+1), z_s^*(k).
\end{align*}

In addition, we consider $ z_s(k+1) = z_s^*(k) $, $ u_s(k+1) = u_s^*(k)$ and $ y_s(k+1) = y_s^*(k) $.

Let us define the shifted error vector:
\begin{align*}
    \bar{e}_z(j) = \bar{z}(j;k+1)-\bar{z}^*(j+1;k), \; j=0,1,\ldots, N-2,
\end{align*}
where $\bar{e}_z(j+1) = A_K \bar{e}_z(j)$ and
\begin{align*}
    \bar{e}_z(0)&=\bar{z}(0;k+1)-\bar{z}^*(1;k)\\
    &= \bar{z}(k+1) - A \bar{z}(k) - B u(k)\\
    &\overset{\eqref{eq:perfect Koopman linear model}}{=} w(k;z,x,u) \in \mathcal{W} := \mathcal{E}(0),
\end{align*}
which implies $\bar{e}_z (j) \in \mathcal{E}(j) = A_K^j \mathcal{W}$, for $j\geq 1$.

Let us iteratively denote the sets $\mathcal{R}(j) := \bigoplus_{i=0}^{j-1} A_K^i \mathcal{W} $, for any $j\geq 1$. It is easy to verify
\begin{align}
    \mathcal{R}(j+1) \ominus \mathcal{E}(j) = \mathcal{R}(j), j\geq 1.
\end{align}

Then, it can be verified that all the constraints are feasible with the above defined input sequence at time step $k+1$.
\begin{itemize}
    \item \textit{State Constraints:} for $j=0,1,\ldots,N-1$, we have
        \begin{align*}
            C_x \bar{z}(j;k+1) = C_x \bar{z}^*(j+1;k) + C_x \bar{e}_z(j).
        \end{align*}
        wherein for $j=0$, it comes
        \begin{align*}
            C_x \bar{z}(0;k+1) &= C_x \bar{z}^*(1;k) + C_x \bar{e}_z(0)\\
            & \in \tilde{\mathcal{X}}(1) \oplus C_x \mathcal{W}\\
            &= \mathcal{X} \ominus \lbrace C_x \mathcal{W} \oplus \mathcal{V} \rbrace \oplus C_x \mathcal{W}\\
            &= \tilde{\mathcal{X}}(0).
        \end{align*}
        
        For $j=1,\ldots, N-1$, it comes
        \begin{align*}
            C_x \bar{z}(j;k+1) &= C_x \bar{z}^*(j+1;k) + C_x \bar{e}_z(j)\\
            & \in \tilde{\mathcal{X}}(j+1) \oplus C_x A_K^j \mathcal{W}\\
            &= \mathcal{X} \ominus \lbrace C_x \mathcal{R}(j+1) \oplus \mathcal{V} \rbrace \oplus C_x A_K^j\mathcal{W}\\
            &= \mathcal{X} \ominus \lbrace C_x \mathcal{R}(j+1) \ominus C_x A_K^j\mathcal{W} \oplus \mathcal{V} \rbrace \\
            &= \tilde{\mathcal{X}}(j).
        \end{align*}
        
        Note $j=N-1$, the state constraint is also satisfied due to the terminal constraint.
        
    \item \textit{Input Constraints:} for $j=0,1,\ldots,N-2$, it can be verified that
        \begin{align*}
            \bar{u}(j;k+1) &= \bar{u}^*(j+1 ; k) \\
            & \quad + K (\bar{z}(j;k+1)-\bar{z}^*(j+1;k)) \\
            &\in \mathcal{U} \ominus K \mathcal{R}(j+1) \oplus K \mathcal{E}(j)\\
            &= \mathcal{U} \ominus K\mathcal{R}(j) = \tilde{\mathcal{U}}(j),\allowdisplaybreaks
        \end{align*}
        and for $j=N-1$, it can be verified that $ \bar{u}(N-1;k+1) =  u_s^*(k) \in \tilde{\mathcal{U}}(N) \subset \tilde{\mathcal{U}}(N-2)$.
        
    \item \textit{Steady state and input constraints:} Since $x_s$ and $u_s$ are decision variables at time step $k+1$, the constraints can be satisfied.
    \item \textit{Terminal constraint:} it can be verified that
    \begin{align*}
        \bar{z}(N;k+1) &= \bar{z}(N+1;k) \\
        &= A \bar{z}^*(N ; k) + B \kappa_f \left ( \bar{z}^*(N; k), y_t(k) \right)\\
        &= A z_s^* (k) + B u_s^*(k)\\
        &= z_s^*(k) = z_{s}(k+1).
    \end{align*}
\end{itemize}

\textbf{(ii) With different references} $y_t(k+1) \not = y_{t}(k)$:
In this case, we can also define a similar shifted input sequence at time step $k+1$ based on the optimal solutions obtained at time step $k$
\begin{align*}
    \bar{u}(0 ; k+1), \ldots, \bar{u}(N-1 ; k), \kappa_f( \bar{z}^*(N; k), y_s(k+1) ),
\end{align*}
with the associated state sequence
\begin{align*}
    \bar{z}(0 ; k), \ldots, \bar{z}(N-1 ; k+1), \bar{z}(N; k+1).
\end{align*}

According to the optimization formulation \eqref{problem:KTMPC}, it can be seen that $y_s$ is a decision variable. Even though the reference $y_t$ is changed from time step $k$ to $k+1$, a solution of $y_s(k+1) = y_s^*(k)$, $ z_s(k+1) = z_s^*(k) $ and $ u_s(k+1) = u_s^*(k)$ are still feasible at time step $k+1$, following the same proof as case (i), with a higher cost for the offset function $\left \| y_s - y_{t} \right \|_{S}^2$. Therefore, all the constraints are feasible at time step $k+1$. The closed-loop system is recursively feasible from a feasible initial state. \qed

\section{Proof of Lemma \ref{lemma:ys bounds}}\label{appendix:lemma 3}

For notation simplicity, let us draw the attention in the 2D case and denote
\begin{align*}
    a = \left \| y_s - y_{t} \right \|, 
    b = \left \| y_s^*(k) - y_{t} \right \|,
    c = \left \| y_s^*(k) - \tilde{y}_{sr} \right \|,
\end{align*}
and with $ \sigma \in [0,1]$, we also know
\begin{align*}
    \sigma c = \left \| y_s^*(k) - y_s \right \|, (1-\sigma) c = \left \| y_s - \tilde{y}_{sr} \right \|.
\end{align*}

\textbf{(i)} If $y_s^*(k)$, $\tilde{y}_{sr}$ and $y_t$ are not in the same line, then we can observe that
\begin{figure}[thbp]
    \centering
    \includegraphics[width=0.6\hsize]{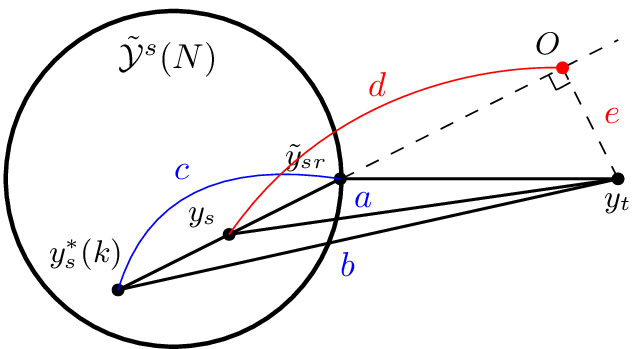}
    \caption{Case (i): the relation of steady output set and $y_s^*(k)$, $\tilde{y}_{sr}$ and $y_t$.}
    \label{fig:ys1}
\end{figure}

As shown in Figure \ref{fig:ys1}, additional two distances $d$ and $e$ are considered with a rectangular vertex $O$. Then, it comes
\begin{align*}
    a^2 &= e^2 + d^2,\\
    b^2 &= e^2 + (d+\sigma c)^2,\\
    d & \geq (1-\sigma) c,
\end{align*}
which implies that
\begin{align*}
    a^2 - b^2 &= d^2 - d^2 - c^2 - 2\sigma cd\\
    & \leq -c^2 - 2\sigma (1-\sigma) c^2\\
    & = -(2\sigma - \sigma^2) c^2.
\end{align*}

\textbf{(ii)} If $y_s^*(k)$, $\tilde{y}_{sr}$ and $y_t$ are in the same line, then we can observe that
\begin{figure}[thbp]
    \centering
    \includegraphics[width=0.6\hsize]{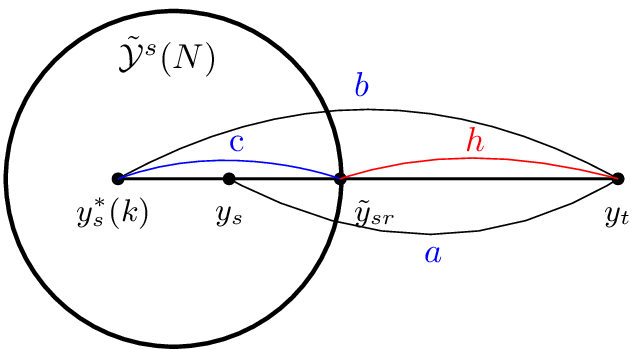}
    \caption{Case (ii): the relation of steady output set and $y_s^*(k)$, $\tilde{y}_{sr}$ and $y_t$.}
    \label{fig:ys2}
\end{figure}

As shown in Figure \ref{fig:ys2}, we have
\begin{align*}
    a &= (1-\sigma)c + h,\\
    b &= c+h.
\end{align*}

Then, we can have 
\begin{align*}
    a^2-b^2 = & (1-\sigma)c^2 + h^2 + 2(1-\sigma)ch \\
    & - c^2 - h^2 - 2 ch\\
    =& -(2\sigma-\sigma^2) c^2 - 2\sigma ch\\
    \leq & -(2\sigma-\sigma^2) c^2.
\end{align*}

Together with cases (i) and (ii), we can therefore obtain that
\begin{align*}
    & \left \| y_s - y_{t} \right \|_{S}^2 - \left \| y_s^*(k) - y_{t} \right \|_{S}^2\\
    =& s \left( \left \| y_s - y_{t} \right \|^2 - \left \| y_s^*(k) - y_{t} \right \|^2 \right)\\
    \leq & -s(2\sigma-\sigma^2) c^2,
\end{align*}
which gives \eqref{eq:y_s(k+1)-ys(k)-yt upper bound}. \qed


\section{Proof of Theorem \ref{theorem:ISS stability}}\label{appendix:theorem-stability}

\textbf{(i) Input-to-State stability to a neighborhood of $y^*_{s}(k)$.} We first prove that the closed-loop state is converging to a neighborhood of optimal reachable steady output $y^*_s(k)$.
    
According to the first Lyapunov candidate function in \eqref{eq:Lyapunov candidate function}, we can have
\begin{align*}
    & \Delta V_1(k;y_t) = V_1(x(k+1),y_t) - V_1(x(k),y_t)\\
    \leq & \sum_{j=0}^{N-1} \ell \left ( \bar{z}(j;k+1)-z_s^*(k), \bar{u}(j;k+1)-u_s^*(k) \right )\\
    & - \sum_{j=0}^{N-1} \ell \left ( \bar{z}^*(j;k)-z_s^*(k), \bar{u}^*(j;k)-u_s^*(k) \right ).
\end{align*}

Let us recall the feasible sequence defined in the proof of Theorem \ref{theorem:RF}. Therefore, we can have
\begin{align*}
    & \sum_{j=0}^{N-1} \left ( \left \|  \bar{z}(j;k+1)-z_s^*(k) \right \|_{Q}^2 + \left \| \bar{u}(j;k+1)-u_s^*(k) \right \|_{R}^2\right )\\
    & \; - \sum_{j=0}^{N-1} \left ( \left \|  \bar{z}^*(j;k)-z_s^*(k) \right \|_{Q}^2 + \left \| \bar{u}^*(j;k)-u_s^*(k) \right \|_{R}^2\right )\allowdisplaybreaks\\
    =& \sum_{j=0}^{N-2} \left ( \left \|  \bar{z}(j;k+1)-z_s^*(k) \right \|_{Q}^2 - \left \|  \bar{z}^*(j+1;k)-z_s^*(k) \right \|_{Q}^2\right )\allowdisplaybreaks\\
    & \; + \sum_{j=0}^{N-2} \left ( \left \| \bar{u}(j;k+1)-u_s^*(k) \right \|_{R}^2 -  \left \| \bar{u}^*(j+1;k)-u_s^*(k) \right \|_{R}^2 \right )\allowdisplaybreaks\\
    & \; + \left \|  \bar{z}(N-1;k+1)-z_s^*(k) \right \|_{Q}^2 \allowdisplaybreaks\\
    & \; + \left \| \bar{u}(N-1;k+1)-u_s^*(k) \right \|_{R}^2\allowdisplaybreaks\\
    & \; - \left \|  \bar{z}^*(0;k)-z_s^*(k) \right \|_{Q}^2 - \left \| \bar{u}^*(0;k)-u_s^*(k) \right \|_{R}^2.
\end{align*}
    
Since we know the error dynamics $ \bar{e}_z(j+1) = A_K \bar{e}_z(j) = A_K^j w(k;z,x,u) $, with $\bar{e}_z(j) = \bar{z}(j;k+1)-\bar{z}^*(j+1;k)$ and $\bar{e}_z(0) = w(k;z,x,u) \in \mathcal{W}$. Based on \eqref{eq:inequality two norm minus}, we can have
\begin{enumerate}
    \item For any $j = 0, \ldots, N-2$, it comes
    \begin{align*}
        & \left \|  \bar{z}(j;k+1)-z_s^*(k) \right \|_{Q}^2 - \left \|  \bar{z}^*(j+1;k)-z_s^*(k) \right \|_{Q}^2\\
        \leq &  \|  A_K^j w(k;z,x,u) \|_{Q}^2 \\
        & \; + 2 \|  A_K^j w(k;z,x,u) \|_{Q} \left \|  \bar{z}^*(j+1;k)-z_s^*(k) \right \|_{Q}\\
        \leq & \|  A_K^j w(k;z,x,u) \|_Q^2 \\
        &\; + 2 \sqrt{J_{N}(x(k),y_t)}\| A_K^j w(k;z,x,u) \|_{Q}.
    \end{align*}
    \item For any $j = 0, \ldots, N-2$, it comes
    \begin{align*}
        & \left \| \bar{u}(j;k+1)-u_s^*(k) \right \|_{R}^2 -  \left \| \bar{u}^*(j+1;k)-u_s^*(k) \right \|_{R}^2\\
        \leq &  \|  A_K^j w(k;z,x,u) \|_{K^{\top}RK}^2 \\
        & \; + 2 \|  A_K^j w(k;z,x,u) \|_{K^{\top}RK} \left \|  \bar{u}^*(j+1;k)-u_s^*(k) \right \|_{R}\\
        \leq & \|  A_K^j w(k;z,x,u) \|_{K^{\top}RK}^2 \\
        & \; + 2 \sqrt{J_{N}(x(k),y_t)} \|  A_K^j w(k;z,x,u) \|_{K^{\top}RK}.
    \end{align*}
    \item Since $\bar{u}(N-1;k+1)=u_s^*(k)$ and $\bar{z}^*(N;k) = z_s^*(k)$, it holds
    \begin{align*}
        &\left \|  \bar{z}(N-1;k+1)-z_s^*(k) \right \|_{Q}^2 \\
        &\; + \left \| \bar{u}(N-1;k+1)-u_s^*(k) \right \|_{R}^2 \\
        =&  \left \|  \bar{z}(N-1;k+1)-\bar{z}^*(N;k) \right \|_{Q}^2\\
        =&  \|  A_K^{N-1} w(k;z,x,u) \|_{Q}^2.
    \end{align*}
\end{enumerate}

Then, we can know that
\begin{align*}
    & \Delta V_1(k;y_t)\\
    \leq & \sum_{j=0}^{N-2} \left( \|  A_K^j w(k;z,x,u) \|_{Q}^2 + \|  A_K^j w(k;z,x,u) \|_{K^{\top}RK}^2 \right)\\
    &\;+ 2 \sqrt{J_{N}(x(k),y_t)} \sum_{j=0}^{N-2} \| A_K^j w(k;z,x,u) \|_{Q} \\
    &\;+ 2 \sqrt{J_{N}(x(k),y_t)} \sum_{j=0}^{N-2} \|  A_K^j w(k;z,x,u) \|_{K^{\top}RK} \\
    &\;+ \|  A_K^{N-1} w(k;z,x,u) \|_{Q}^2 \\
    &\;- \left \|  \bar{z}^*(0;k)-z_s^*(k) \right \|_{Q}^2 - \left \| \bar{u}^*(0;k)-u_s^*(k) \right \|_{R}^2.
\end{align*}

Let us denote
\begin{align*}
    \gamma(w) := & \sum_{j=0}^{N-2} \|  A_K^j w(k;z,x,u) \|_{Q}^2 \\
    &\; + \sum_{j=0}^{N-2}\|  A_K^j w(k;z,x,u) \|_{K^{\top}RK}^2 \\
    &\;+ 2 \sqrt{J_{N}(x(k),y_t)} \sum_{j=0}^{N-2} \| A_K^j w(k;z,x,u) \|_{Q} \\
    &\;+ 2 \sqrt{J_{N}(x(k),y_t)} \sum_{j=0}^{N-2} \|  A_K^j w(k;z,x,u) \|_{K^{\top}RK} \\
    &\;+ \|  A_K^{N-1} w(k;z,x,u) \|_{Q}^2,
\end{align*}
which is a $\mathcal{K}_{\infty}$ function. Then, the above condition can be simplified as
\begin{align*}
    \Delta V_1(k;y_t)
    \leq &  - \left \|  \bar{z}^*(0;k)-z_s^*(k) \right \|_{Q}^2 - \left \| \bar{u}^*(0;k)-u_s^*(k) \right \|_{R}^2\\
    & \; + \gamma(w) \\
    \leq &  - \left \|  \bar{z}^*(0;k)-z_s^*(k) \right \|_{Q}^2 + \gamma(w).
\end{align*}

Therefore, with \eqref{eq:KTMPC initialization}, we can obtain
\begin{align}\label{eq:Delta V1 upper bound}
    \Delta V_1(k;y_t) &\leq -\left \| z(k) - z_s^* (k) \right\|_Q^2 + \gamma(w) \nonumber\\
    & \overset{\eqref{eq:V1 bounds}}{\leq} - \beta_{l_1} \left \| z(k) - z_s^* (k) \right\|^2 + \gamma(w),
\end{align}
which indicates that the closed-loop system is input-to-state stable to a neighborhood of $y_s^*(k)$.

From \eqref{eq:V1 bounds}, we also know a lower bound of $\Delta V_1(k;y_t)$ as
\begin{align}\label{eq:Delta V1 lower bound}
    \Delta V_1(k;y_t) \geq & \; \beta_{l_1}\left \| z(k+1) - z_s^* (k+1) \right\|^2 \nonumber\\
    & - \beta_{u_1}\left \| z(k) - z_s^* (k) \right\|^2,
\end{align}
and together with \eqref{eq:Delta V1 upper bound} and \eqref{eq:Delta V1 lower bound}, we have
\begin{align}\label{eq:zk-zs iss}
    \left \| z(k+1) - z_s^* (k+1) \right\|^2 \leq & \; c_1 \left \| z(k) - z_s^* (k) \right\|^2 + \frac{1}{\beta_{l_1}}\gamma(w),
\end{align}
where $c_1 := \frac{\beta_{u_1}-\beta_{l_1}}{\beta_{l_1}}$ with $ 0 < c_1 < 1 $ due to \eqref{eq:beta u1 condition}, which can also provide practical exponential stability for bounded $w$.



\textbf{(ii) Input-to-State stability to a neighborhood of $\tilde{y}_{sr}$.} We next discuss that the closed-loop system output $y(k)$ is converging to a neighborhood of $\tilde{y}_{sr}$. Here, we rely on the second Lyapunov candidate function defined in \eqref{eq:Lyapunov candidate function 2}.
\begin{align*}
    \Delta V_2(k;y_t) = &V_2(x(k+1),y_t) - V_2(x(k),y_t)\\
    = & J_{N}^{*}(x(k+1),y_t) - J_{N}^{*}(x(k),y_t).
\end{align*}


To discuss the bound for $V_2(x(k+1),y_t) - V_2(x(k),y_t)$, we consider the following two complementary cases:
\begin{enumerate}
    \item[a)] There exists a scalar $\rho > 0$ such that
    \begin{align}\label{eq:case 1 condition}
        \left \|  z(k)-z^*_s(k) \right \|^2 \geq \rho \left \|  z^*_s(k) - \tilde{z}_{sr} \right \|^2.
    \end{align}
    \item[b)] There exists a scalar $\rho > 0$ such that
    \begin{align}\label{eq:case 2 condition}
        \left \|  z(k)-z^*_s(k) \right \|^2 \leq \rho \left \|  z^*_s(k) - \tilde{z}_{sr} \right \|^2.
    \end{align}
\end{enumerate}

The above two cases are considered in order to assign feasible solutions for KTMPC at time step $k+1$. Since the same $\rho$ is used in \eqref{eq:case 1 condition} and \eqref{eq:case 2 condition}, the closed-loop stability result does not depend on any of these two conditions.

\textbf{Case a):} In this case, we can choose the same non-optimal solutions at time step $k+1$ as in Part (i) and it comes
\begin{align*}
    \Delta V_2(k;y_t) 
    \leq &  J_{N}(x(k+1),y_t) - J_{N}^{*}(x(k),y_t)\\
    =& \left \| y_s(k+1) - y_{t} \right \|_{S}^2 - \left \| y_s^*(k) - y_{t} \right \|_{S}^2 \\
    & \; + \sum_{j=0}^{N-1} \left \|  \bar{z}(j;k+1)-z_s(k+1) \right \|_{Q}^2 \\
    & \; + \sum_{j=0}^{N-1}\left \| \bar{u}(j;k+1)-u_s(k+1) \right \|_{R}^2\\
    & \; - \sum_{j=0}^{N-1} \Big ( \left \|  \bar{z}^*(j;k)-z_s^*(k) \right \|_{Q}^2 \\
    & \qquad \qquad + \left \| \bar{u}^*(j;k)-u_s^*(k) \right \|_{R}^2\Big ).
\end{align*}

Then, it can be realized that
\begin{align*}
    & \Delta V_2(k;y_t) \leq \Delta V_1(k;y_t) \overset{\eqref{eq:Delta V1 upper bound}}{\leq} - \left \| z(k) - z_s^* (k) \right\|_Q^2 + \gamma(w).
\end{align*}

Therefore, we can further derive that
\begin{align*}
    & \left \|  z(k)-z^*_s(k) \right \|_Q^2\\
    = & \; \frac{1}{2} \left \|  z(k)-z^*_s(k) \right \|_Q^2 + \frac{1}{2} \left \|  z(k)-z^*_s(k) \right \|_Q^2\\
    \geq & \; \frac{1}{2} \left \|  z(k)-z^*_s(k) \right \|_Q^2  + \frac{\underline{\lambda}(Q)}{2} \left \|  z^*_s(k) - \tilde{z}_{sr} \right \|_Q^2\\
    \,\overset{\eqref{eq:case 1 condition}}{\geq} & \; \frac{\underline{\lambda}(Q)}{2} \left \|  z(k)-z^*_s(k) \right \|^2  + \frac{\rho \underline{\lambda}(Q)}{2} \left \|  z^*_s(k) - \tilde{z}_{sr} \right \|^2\\
    \geq & \;\tilde{c}_1 \left \|  z(k) - \tilde{z}_{sr} \right \|^2,
\end{align*}
where $\tilde{c}_1 := \frac{\min\{\underline{\lambda}(Q),\rho  \underline{\lambda}(Q)\}}{4} $ with $\tilde{c}_1>0$ due to $Q \in \mathbb{S}^{n_x}_{\succ 0}$.
  
Thus, we can conclude for this case
\begin{align}\label{eq:Delta V2 upper bound 1}
    \Delta V_2(k;y_t) \leq - \tilde{c}_1  \left \|  z(k) - \tilde{z}_{sr} \right \|^2 + \gamma(w),
\end{align}
which guarantees that the closed-loop state $z(k)$ is input-to-state stable and converging to a neighbourhood of $\tilde{z}_{sr}$.

\textbf{Case b):} Given the optimal solutions obtained at time step $k$, we can consider a non-optimal solution of steady output at the next time step $k+1$ as
\begin{align*}
    y_s(k+1) &= \sigma y_s^*(k) + (1-\sigma) \tilde{y}_{sr}, \; \sigma \in (0,1],
\end{align*}
which can imply non-optimal steady state $z_s(k+1)$ and input $u_s(k+1)$.

Then, we can have that there exists a scalar $\beta_2 > 0 $ such that
\begin{align*}
    \Delta V_2(k;y_t) \leq & V_1(x(k+1),y_t) + \left \| y_s(k+1) - y_{t} \right \|_{S}^2 \\
    & - V_1(x(k),y_t) - \left \| y_s^*(k) - y_{t} \right \|_{S}^2 \\
    \overset{\eqref{eq:V1 bounds}}{\leq} & \left \| y_s(k+1) - y_{t} \right \|_{S}^2 - \left \| y_s^*(k) - y_{t} \right \|_{S}^2\\
    & \; + \beta_{u_1} \left \|  z(k+1)-z_s^*(k+1) \right \|^2 \\
    & \; - \left \|  z(k) - z_s^*(k) \right \|_{Q}^2.
\end{align*}
    
First, we can derive the term
\begin{align*}
    & \left \| y_s(k+1) - y_{t} \right \|_{S}^2 - \left \| y_s^*(k) - y_{t} \right \|_{S}^2\\
    \overset{\eqref{eq:y_s(k+1)-ys(k)-yt upper bound}}{\leq} & \; -s(2\sigma-\sigma^2) \left \| y_s^*(k) - \tilde{y}_{sr} \right \|^2\\
    \overset{\eqref{eq:ys lower bound}}{\leq} &\; -\frac{s(2\sigma-\sigma^2)}{c_l}\left \| z_s^*(k) - \tilde{z}_{sr} \right \|^2.
\end{align*}

Second, from \eqref{eq:zk-zs iss}, we can know
\begin{align*}
    & \left \|  z(k+1)-z_s^*(k+1) \right \|^2 \leq c_1 \left \| z(k) - z_s^* (k) \right\|^2 + \frac{1}{\beta_{l_1}} \gamma(w),
\end{align*}

Then, by combining the above inequality conditions, we can obtain that
\begin{align*}
    \Delta V_2(k;y_t) \leq & \; - \beta_{l_1} \left \|  z(k) - z_s^*(k) \right \|^2\\
    & \;-\frac{s(2\sigma-\sigma^2)}{c_l}\left \| z_s^*(k) - \tilde{z}_{sr} \right \|^2 \\
    & \;+ \beta_{u_1} c_1 \left \| z(k)- z_s^*(k \right \|^2 + \frac{\beta_{u_1}}{\beta_{l_1}} \gamma(w).
\end{align*}

Let us denote $ \tilde{\gamma}(w) = \frac{\beta_{u_1}}{\beta_{l_1}} \gamma(w) $ and $ c_2 = \frac{s(2\sigma-\sigma^2)}{c_l} > 0 $ (due to $s > 0 $ and $0<\sigma <1$). Therefore, the above condition can be simplified as
\begin{align}\label{eq:Delta V2 upper bound 2}
    \Delta V_2(k;y_t) \leq & - (\beta_{l_1}-\beta_{u_1} c_1)\left \|  z(k) - z_s^*(k) \right \|^2  \\
    & - c_2 \left \| z_s^*(k) - \tilde{z}_{sr} \right \|^2 +  \tilde{\gamma}(w)\nonumber \\
    \overset{\eqref{eq:inequality sum norm bound}}{\leq} & - \tilde{c}_2 \left \|  z(k) - \tilde{z}_{sr} \right \|^2 + \tilde{\gamma}(w).
\end{align}
where $\tilde{c}_2 := \frac{\min\{(\beta_{l_1}-\beta_{u_1} c_1),c_2\}}{2} $. From Assumption \ref{assump:beta u bound}, the condition \eqref{eq:beta u1 condition} gives 
\begin{align*}
    \beta_{u_1}^2 - \beta_{u_1} \beta_{l_1} -\beta_{l_1}^2 <0,
\end{align*}
which ensure the following condition with $ \beta_{l_1} =\underline{\lambda}(Q) > 0 $,
\begin{align*}
    \beta_{l_1}- \frac{\beta_{u_1}(\beta_{u_1}-\beta_{l_1})}{\beta_{l_1}} > 0,
\end{align*}
which is equivalent to $\beta_{l_1}-\beta_{u_1} c_1 > 0$. Thus, $\tilde{c}_2 > 0$ holds together with $c_2 > 0$.

Together with Lemma \ref{lemma:bounds for V2} and conditions in \eqref{eq:Delta V2 upper bound 1} and \eqref{eq:Delta V2 upper bound 2}, we can concluded that the closed-loop system state $z(k)$ is converging to a neighborhood of $\tilde{z}_{sr}$, which implies the closed-loop system output $ y(k) $ of \eqref{eq:general nonlinear system} with the KTMPC controller defined in~\eqref{problem:KTMPC} is converging to a neighborhood of the given reference $y_t$. In addition, from the obtained Lyapunov conditions above, it can be seen that the size of the neighborhood is determined by the amplitude of estimation errors. \qed


\ifCLASSOPTIONcaptionsoff
  \newpage
\fi



\bibliographystyle{IEEEtran}
\bibliography{IEEEabrv,KTMPC}

\end{document}